%% file: main.tex
\documentclass[copyright,creativecommons]{eptcs}
\usepackage{latex/packages}            
\input{latex/beluga-listings}              
\input{latex/macro}   			
\usepackage{latex/macros-arrows}	
\graphicspath{{figures}}

\title{A Formalization of the Reversible

Concurrent Calculus \pccsk in Beluga}
\author{Gabriele Cecilia
	\orcidlink{0009-0007-7797-5008}
\institute{School of Computer \& Cyber Sciences,\\
 Augusta University, Augusta, USA\\
	\email{gcecilia@augusta.edu}
}}
\def\titlerunning{A Formalization of the Reversible Concurrent Calculus \pccsk in Beluga}
\def\authorrunning{G.\ Cecilia}

\begin{document}
\maketitle

\begin{abstract}
Reversible concurrent calculi are abstract models for concurrent systems in which any action can potentially be undone. Over the last few decades, different formalisms have been developed and their mathematical properties have been explored; however, none have been machine-checked within a proof assistant. 
This paper presents the first Beluga formalization of the Calculus of Communicating Systems with Keys and Proof labels (\pccsk), a reversible extension of \ccs. 
Beyond the syntax and semantics of the calculus, the encoding covers state-of-the-art results regarding three relations over proof labels -- namely, dependence, independence and connectivity -- which offer new insights into the notions of causality and concurrency of events. 
As is often the case with formalizations, our encoding introduces adjustments to the informal proof and makes explicit details which were previously only sketched, some of which reveal to be less straightforward than initially assumed. 
We believe this work lays the foundations for future reversible concurrent calculi formalizations.
\end{abstract}

\input{sections/introduction.tex}

\input{sections/ccskp.tex}

\input{sections/beluga.tex}

\input{sections/conclusions.tex}

\section*{Acknowledgments}
	We warmly thank the anonymous reviewers, as well as Clément Aubert and Deivid Vale, for their insightful comments and suggestions, which greatly contributed to improving the paper.
	This work is supported by the National Science Foundation under \href{https://www.nsf.gov/awardsearch/showAward?AWD_ID=2242786}{Grant No. 2242786 (SHF:Small:Concurrency In Reversible Computations)}.	
	
\bibliographystyle{eptcs}
\bibliography{bib/bib}

\end{document}

%% file: latex/beluga-listings.tex
\newcommand*{\emacsfont}{\fontfamily{lmtt}\selectfont}

\definecolor{belugapurple}{RGB}{154,0,214}
\definecolor{belugared}{RGB}{225, 0, 0}
\definecolor{belugagreen}{RGB}{31,147,15}
\definecolor{belugablue}{RGB}{0,0,238}
\definecolor{belugapink}{RGB}{237,103,166}
\newcommand\delimone{{\color{belugapurple}\emacsfont\bfseries LF}\color{belugagreen}\aftergroup:}
\newcommand\delimtwo{{\color{belugapurple}\emacsfont\bfseries inductive}\color{belugagreen}\aftergroup:}
\newcommand\delimthree{{\color{belugapurple}\emacsfont\bfseries schema}\color{belugagreen}\aftergroup=}
\newcommand\delimfour{{\color{belugapurple}\emacsfont\bfseries rec}\color{belugablue}\aftergroup:}
\newcommand\delimfive{{/}\slshape\aftergroup/}
\newcommand\delimaux{{\color{belugapink}.}}
\newcommand\delimsix{\color{belugapink}--\aftergroup\delimaux}

\lstdefinelanguage{Beluga}
{
  morekeywords=[1]{mlam,fn,case,of,total,in,type, impossible, let, and,ctype},
  keywordstyle=[1]\color{belugapurple}\emacsfont\bfseries,
  morecomment=[l]{\%},
  morecomment=[s]{\%\{}{\}\%},
  commentstyle=\color{belugared},
  sensitive=true,
}

\lstdefinestyle{belugastyle}{
 language=Beluga,
    basicstyle=\ttfamily\small,
    columns=flexible,
    keepspaces=true,
    showstringspaces=false,
    breaklines=true,
    breakatwhitespace=true,
    rangeprefix=\%\%\ ,
    rangesuffix=\ \%\%,
    includerangemarker=false,
    moredelim=**[is][\delimone]{LF}{:},
    moredelim=**[is][\delimtwo]{inductive}{:},
    moredelim=**[is][\delimthree]{schema}{=},
    moredelim=**[is][\delimfour]{rec}{:},
    moredelim=**[is][\delimfive]{/}{/},
    moredelim=**[is][\delimsix]{--}{.},
    literate={→}{{$\rightarrow$}}{1}
                  {⊢}{{$\vdash$}}{1}
                  {⇒}{{$\Rightarrow$}}{1}
}

\lstdefinestyle{belugastyleframes}{
 language=Beluga,
    basicstyle=\ttfamily\small,
    columns=flexible,
    keepspaces=true,
    showstringspaces=false,
    breaklines=true,
    breakatwhitespace=true,
    rangeprefix=\%\%\ ,
    rangesuffix=\ \%\%,
    includerangemarker=false,
    moredelim=**[is][\delimone]{LF}{:},
    moredelim=**[is][\delimtwo]{inductive}{:},
    moredelim=**[is][\delimthree]{schema}{=},
    moredelim=**[is][\delimfour]{rec}{:},
    moredelim=**[is][\delimfive]{/}{/},
    moredelim=**[is][\delimsix]{--}{.},
    frame=tblr,
    captionpos=b,
    literate={→}{{$\rightarrow$}}{1}
                  {⊢}{{$\vdash$}}{1}
                  {⇒}{{$\Rightarrow$}}{1}
}

%% file: latex/macro.tex
\usepackage{ushort}

\newcommand{\eveqt}{\sim}

\newcommand{\Par}{\mid}

\newcommand{\conn}{\mathrel{\curlyvee}} 

\newcommand{\rev}[1]{\ushortw{#1}} 
\newcommand{\Rev}[1]{\ensuremath{\rev{#1}}} 
\newcommand{\tRev}[1]{\Rev{\text{#1}}}

\newcommand{\ind}{\mathrel{\iota}}

\makeatletter
\newcommand*{\etc}{%
	\@ifnextchar{.}%
	{etc}
	{etc.\@\xspace}%
}
\makeatother

\makeatletter
\DeclareFontEncoding{LS2}{}{\@noaccents}
\makeatother
\DeclareFontSubstitution{LS2}{stix}{m}{n}
\DeclareSymbolFont{largesymbolsstix}{LS2}{stixex}{m}{n}
\DeclareMathDelimiter{\lBrace}{\mathopen} {largesymbolsstix}{"E8}{largesymbolsstix}{"0E}
\DeclareMathDelimiter{\rBrace}{\mathclose}{largesymbolsstix}{"E9}{largesymbolsstix}{"0F}

\DeclareFontEncoding{LS1}{}{}
\DeclareFontSubstitution{LS1}{stix}{m}{n}
\DeclareSymbolFont{stixsymbols}{LS1}{stixscr}{m}{n}
\SetSymbolFont{stixsymbols}{bold}{LS1}{stixscr}{b}{n}

\makeatletter
\NewDocumentCommand{\smallbar}{}{%
	\mathrel{\mathpalette\smallbar@\relax}%
}

\newcommand{\current@math@font}[1]{%
	\ifx#1\displaystyle\textfont\else
	\ifx#1\textstyle\textfont\else
	\ifx#1\scriptstyle\scriptfont\else
	\scriptscriptfont\fi\fi\fi
}
\newcommand{\smallbar@factor}[1]{%
	\ifx#1\displaystyle 1.135\else
	\ifx#1\textstyle 1.128\else
	\ifx#1\scriptstyle 1.09\else
	1.06\fi\fi\fi
}

\newcommand{\smallbar@}[2]{%
	\begingroup
	\sbox\z@{$\m@th#1\mapstochar$}%
	\dimen0=\smallbar@factor{#1}\ht\z@
	\dimen2=\dimeval{2\fontdimen22\current@math@font{#1} 2 - \dimen0}%
	\mbox{%
		$\m@th#1\mkern1mu
		\begin{picture}(0,\dimen0)
			\roundcap
			\linethickness{\fontdimen8\current@math@font{#1}3}
			\Line(0,\dimen2)(0,\dimen0)
		\end{picture}%
		\mkern1mu$%
	}%
	\endgroup
}
\makeatother

\newcommand{\orig}[1]{O_{#1}} 
\newcommand{\real}[1]{r(#1)} 
\newcommand{\bs}{\backslash}

\DeclareSymbolFont{symbolsC}{U}{txsyc}{m}{n}
\SetSymbolFont{symbolsC}{bold}{U}{txsyc}{bx}{n}
\DeclareFontSubstitution{U}{txsyc}{m}{n}
\DeclareMathSymbol{\opentimes}{\mathrel}{symbolsC}{93}
\newcommand{\sdep}{\opentimes}

\newcommand{\Left}{\mathrm{L}}        
\newcommand{\Right}{\mathrm{R}}       
\newcommand{\Dir}{\mathrm{d}}         
\newcommand{\OpDir}[1]{\overline{#1}} 

\renewcommand{\L}{\Left}              
\newcommand{\R}{\Right}
\newcommand{\D}{\Dir}
\newcommand{\OD}{\OpDir{\D}}

\newcommand{\lmidr}{{\mid_{\R}}}
\newcommand{\lmidl}{{\mid_{\L}}}
\newcommand{\lmidd}{{\mid_{\D}}}
\newcommand{\lmidod}{{\mid_{\OD}}}

\newcommand{\lplusl}{{+_{\L}}}
\newcommand{\lplusd}{{+_{\D}}}
\newcommand{\lplusod}{{+_{\OD}}}

\ushortCreate()[.05em](1.2){ushortwt} 

\newcommand{\rlmidl}{{\ushortwt{\mid_{\L}}}}

\newcommand{\rlplusl}{{\Rev{+_{\L}}}}

\newcommand{\bpair}[2]{\langle #1 , #2 \rangle} 
\newcommand{\cpair}[2]{\bpair{\lmidl #1}{\lmidr #2}} 

\newcommand{\names}{\ensuremath{\mathsf{N}}}

\newcommand{\labelset}{\ensuremath{\mathsf{L}}}
\newcommand{\keyset}{\ensuremath{\mathsf{K}}}
\newcommand{\proofset}{\ensuremath{\mathsf{P}}}
\newcommand{\kplabelset}{\labelset_{\keyset}^{\proofset}}

\newcommand{\out}[1]{\overline{#1}}

\newcommand{\ccs}{CCS\xspace}
\newcommand{\ccsk}{CCSK\xspace}
\newcommand{\rccs}{RCCS\xspace}

\newcommand{\pccsk}{CCSK$^{\text{P}}$\xspace}

\DeclareMathOperator{\stdop}{\mathsf{std}}

\DeclareMathOperator{\keysop}{\mathsf{keys}}

\DeclareMathOperator{\lablop}{\ell} 
\DeclareMathSymbol{\kayop}{\mathalpha}{stixsymbols}{"6B} 
\DeclareMathSymbol{\ekayop}{\mathalpha}{stixsymbols}{"62} 

\newcommand{\std}[1]{\stdop(#1)}

\newcommand{\kay}[1]{\kayop(#1)}

\newcommand{\keys}[1]{\keysop(#1)}

\newcommand{\labl}[1]{\lablop(#1)}

\ExplSyntaxOn
\NewDocumentCommand{\Rel}{m O{} O{}} 
{
	\str_case:nnF { #1 }
	{
		{s}{ 
			\tl_if_blank:nTF {#2} {       							
				\tl_if_blank:nTF {#3} {	  							
					\mathrel{<} 	    				
				}{
					\mathrel{<^{#3}}					
			}}{
				\tl_if_blank:nTF {#3} {	  							
					\mathrel{<\c_math_subscript_token{\scriptscriptstyle{\mathsf{#2}}}} 		
				}{
					\mathrel{<\c_math_subscript_token{\scriptscriptstyle{\mathsf{#2}}}^{#3}} 	
			}}
		}
		{i}{ 
			\tl_if_blank:nTF {#2} {       							
				\tl_if_blank:nTF {#3} {	  							
					\mathrel{>} 	    				
				}{
					\mathrel{>^{#3}}					
			}}{
				\tl_if_blank:nTF {#3} {	  							
					\mathrel{>\c_math_subscript_token{\scriptscriptstyle{\mathsf{#2}}}} 		
				}{
					\mathrel{>\c_math_subscript_token{\scriptscriptstyle{\mathsf{#2}}}^{#3}} 	
			}}
		}	
		{b}{ 
			\tl_if_blank:nTF {#2} {       							
				\tl_if_blank:nTF {#3} {	  							
					\mathrel{\eveqt} 	    				
				}{
					\mathrel{\eveqt^{#3}}					
			}}{
				\tl_if_blank:nTF {#3} {	  							
					\mathrel{\eveqt\c_math_subscript_token{\scriptscriptstyle{\mathsf{#2}}}} 		
				}{
					\mathrel{\eveqt\c_math_subscript_token{\scriptscriptstyle{\mathsf{#2}}}^{#3}} 	
			}}
		}
		{r}{ 
			\tl_if_blank:nTF {#2} {       							
				\tl_if_blank:nTF {#3} {	  							
					\mathrel{\mathcal{R}} 	    				
				}{
					\mathrel{\mathcal{R}^{#3}}					
			}}{
				\tl_if_blank:nTF {#3} {	  							
						\mathcal{R}\c_math_subscript_token{\scriptscriptstyle{\mathsf{#2}}}
				}{
						\mathcal{R}\c_math_subscript_token{\scriptscriptstyle{\mathsf{#2}}}^{#3}
			}}
		}
	}{error}
}
\NewDocumentCommand{\nRel}{m O{} O{}} 
{
	\str_case:nnF { #1 }
	{
		{s}{ 
			\tl_if_blank:nTF {#2} {       							
				\tl_if_blank:nTF {#3} {	  							
					\mathrel{\not{<}} 	    				
				}{
					\mathrel{\not{<^{#3}}}					
			}}{
				\tl_if_blank:nTF {#3} {	  							
					\mathrel{\not{<\c_math_subscript_token{\scriptscriptstyle{\mathsf{#2}}}}} 		
				}{
					\mathrel{\not{<\c_math_subscript_token{\scriptscriptstyle{\mathsf{#2}}}^{#3}}} 	
			}}
		}
		{i}{ 
			\tl_if_blank:nTF {#2} {       							
				\tl_if_blank:nTF {#3} {	  							
					\mathrel{\not{>}} 	    				
				}{
					\mathrel{\not{>^{#3}}}					
			}}{
				\tl_if_blank:nTF {#3} {	  							
					\mathrel{\not{>\c_math_subscript_token{\scriptscriptstyle{\mathsf{#2}}}}} 		
				}{
					\mathrel{\not{>\c_math_subscript_token{\scriptscriptstyle{\mathsf{#2}}}^{#3}}} 	
			}}
		}	
		{b}{ 
			\tl_if_blank:nTF {#2} {       							
				\tl_if_blank:nTF {#3} {	  							
					\mathrel{\nsim} 	    				
				}{
					\mathrel{\nsim^{#3}}					
			}}{
				\tl_if_blank:nTF {#3} {	  							
					\mathrel{\nsim\c_math_subscript_token{\scriptscriptstyle{\mathsf{#2}}}} 		
				}{
					\mathrel{\nsim\c_math_subscript_token{\scriptscriptstyle{\mathsf{#2}}}^{#3}} 	
			}}
		}
	}{error}
}
\ExplSyntaxOff

\definecolor{irek}{HTML}{1b9e77}
\definecolor{iain}{HTML}{d95f02}
\definecolor{clem}{HTML}{7570b3}
\newcommand{\Comment}[1]{}

\definecolor{ind} {RGB}{43,131,186}
\colorlet  {conc}{ind!30}
\definecolor{sdep}{RGB}{215,25,28}
\colorlet  {tsdep}{sdep!30}
\definecolor{conf}{RGB}{171,221,164}
\definecolor{ord} {RGB}{253,174,97}

\tikzset{
sys/.style = {
	line width=.5mm,
	inner sep=0pt,
	outer sep=0pt,
	rounded corners=0.5cm,
	draw = tsdep,
	minimum size=1.5cm,
	text width=1.6cm,
	text centered
},
spl/.style = {
	rectangle split,
	rectangle split horizontal,
	rectangle split parts=2,
	rectangle split part fill={sdep!20,ind!20},
},
sya/.style= {
	thick,
	font = {\LARGE},	
},
rel/.style = {
	line width=1mm, 
},
adj/.style = {
	line width=1mm, 
	dotted,
},
cpt/.style = {
},
u/.style = {yshift=#1\pgflinewidth},
s/.style = {xshift=#1\pgflinewidth},
}

\usetikzlibrary{calc,decorations.pathreplacing}

\makeatletter
\appto{\endmulticols}{\@doendpe}
\makeatother

\newtheorem{theorem}{Theorem}[section]
\newtheorem{lemma}[theorem]{Lemma}
\newtheorem{definition}[theorem]{Definition}
\newtheorem{revtheorem}{Theorem 2.\hspace{-1mm}}
\newcommand{\theoremitemref}[2]{%
\hyperref[#2]{\ref*{#1}\ref*{#2}}%
}
\newtheorem{example}{Example}
\newcommand{\type}[1]{{\ttfamily \color{belugagreen}#1}}
\newcommand{\keyw}[1]{{\ttfamily \color{belugapurple}#1}}
\newcommand{\func}[1]{{\ttfamily \color{belugablue}#1}}
\newcommand{\baselink}[4]{\href{https://github.com/CinRC/A-Beluga-Formalization-of-CCSKP/blob/main/code/#1\#L#2-L#3}{\emph{#4}}}
\newcommand{\linktype}[4]{\href{https://github.com/CinRC/A-Beluga-Formalization-of-CCSKP/blob/main/code/#1\#L#2-L#3}{\type{#4}}}
\newcommand{\linkfun}[4]{\href{https://github.com/CinRC/A-Beluga-Formalization-of-CCSKP/blob/main/code/#1\#L#2-L#3}{\func{#4}}}

\newcommand{\ExternalLink}{%
    \tikz[x=1.2ex, y=1.2ex, baseline=-0.05ex]{%
        \begin{scope}[x=1ex, y=1ex]
            \clip (-0.1,-0.1) 
                --++ (-0, 1.2) 
                --++ (0.6, 0) 
                --++ (0, -0.6) 
                --++ (0.6, 0) 
                --++ (0, -1);
            \path[draw, 
                line width = 0.5, 
                rounded corners=0.5] 
                (0,0) rectangle (1,1);
        \end{scope}
        \path[draw, line width = 0.5] (0.5, 0.5) 
            -- (1, 1);
        \path[draw, line width = 0.5] (0.6, 1) 
            -- (1, 1) -- (1, 0.6);
        }
    }
\newcommand{\extlink}[3]{\href{https://github.com/CinRC/A-Beluga-Formalization-of-CCSKP/blob/main/code/#1\#L#2-L#3}{\ExternalLink}}

\let\oldalpha\alpha
\renewcommand{\alpha}{\ensuremath{\scalebox{0.88}{$\oldalpha$}}}
\let\oldbeta\beta
\renewcommand{\beta}{\ensuremath{\scalebox{0.88}{$\oldbeta$}}}
\let\oldeta\eta
\renewcommand{\eta}{\ensuremath{\scalebox{0.88}{$\oldeta$}}}
\let\oldtheta\theta
\renewcommand{\theta}{\ensuremath{\scalebox{0.88}{$\oldtheta$}}}
\let\oldlambda\lambda
\renewcommand{\lambda}{\ensuremath{\scalebox{0.88}{$\oldlambda$}}}
\let\oldpi\pi
\renewcommand{\pi}{\ensuremath{\scalebox{0.88}{$\oldpi$}}}

%% file: sections/introduction.tex
\section{Introduction}\label{sec:introduction}

Concurrency in computer science refers to the simultaneous execution of multiple operations or computations in a shared environment. It is a fundamental aspect of modern computing, with practical use in several domains such as operating systems, networking and distributed systems.
Process calculi like \ccs~\cite{Milner1980} and the $\pi$-calculus \cite{Milner1992a} are well-studied and established mathematical models for formally describing and reasoning about concurrent systems.

\begin{figure}[ht]
    \centering
    \fbox
    {
        \begin{minipage}{0.48\textwidth} 
            \centering
            \includegraphics[width=\linewidth]{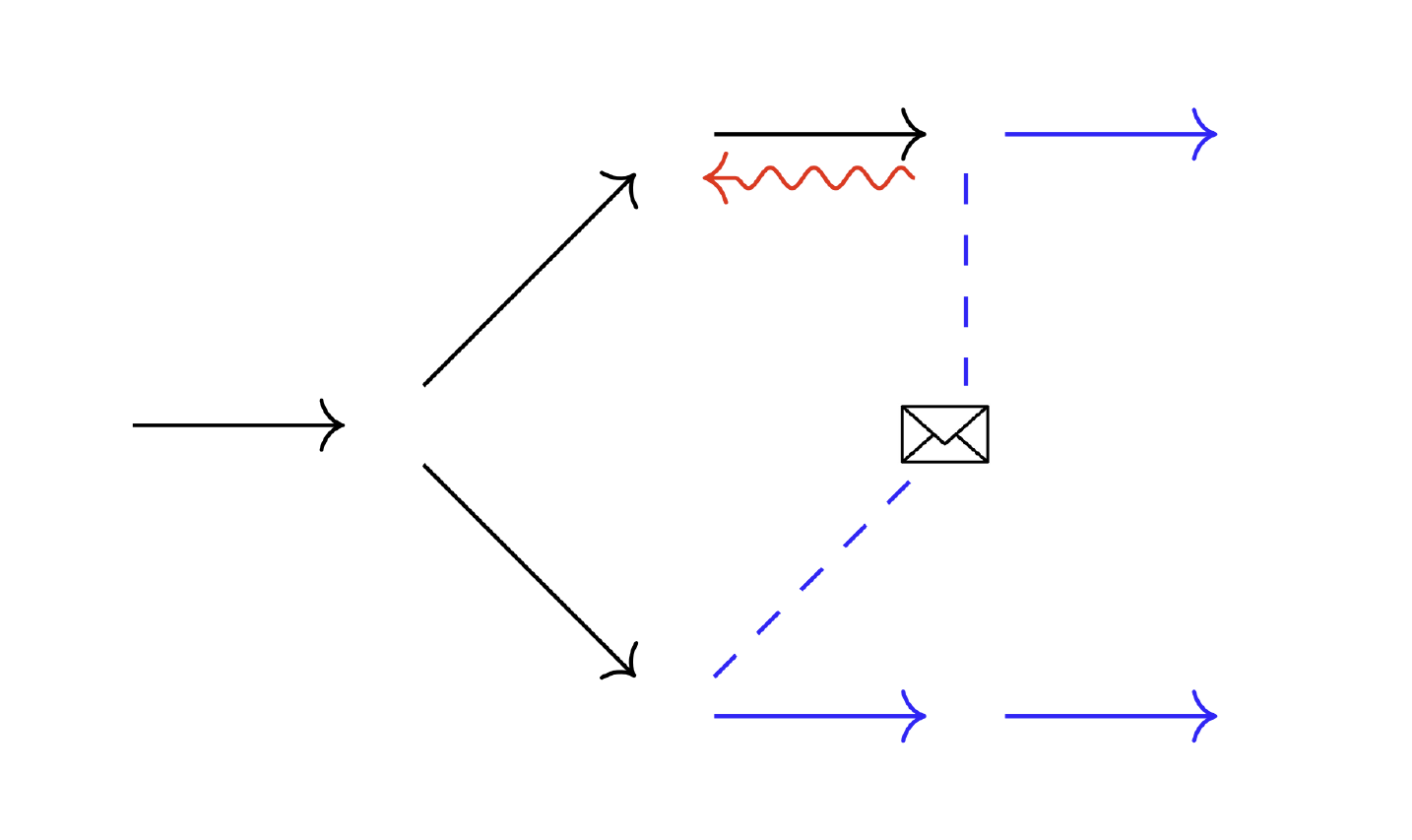}
        \end{minipage}
        \hfill
        \begin{minipage}{0.45\textwidth} 
            \emph{The diagram on the left illustrates the duplication of a thread, followed by a message exchange between the resulting branches and some further independent computation steps. To undo the computation step marked in red, which belongs to one of the two threads and precedes their interaction, it is necessary to first undo each of the steps marked in blue, including those of the other thread.}
        \end{minipage}
    }
    \caption{Example of reversal of computation steps in a concurrent setting.}
    \label{fig:rev-example}
\end{figure}

In recent years, reversing computations in concurrent systems has gained significant attention, with applications in fields like hardware, software and biochemistry \cite{Ulidowski2014}. Enriching concurrent systems with reversibility poses its own set of challenges: for instance, it requires providing some kind of history-preserving mechanism to take track of past actions. Additionally, undoing computation steps in a parallel setting is more complex than in a sequential system: as explained in Fig.~\ref{fig:rev-example}, reversing a specific action performed by a single thread may require knowing, and eventually undoing, the actions of the other threads it has previously interacted with.

Reversible concurrent calculi address such challenges in various ways. 
For example, Reversible \ccs (\rccs) \cite{Danos2004} equips processes with a memory that records information about past computations; conversely, \ccs with Keys (\ccsk) \cite{Phillips2007b} associates unique keys to each forward action. The latter has been recently upgraded to \ccsk with Proof labels (\pccsk) \cite{aubert2022c}, which features a proved transition system in the fashion of Degano and Priami \cite{Degano2001}; proof labels enable the definition of dependence and independence for both forward and backward transitions.
In this framework, the contributions brought by Aubert et al.~\cite{aubert25} merit attention. The authors are the first to introduce separate axioms for the relations of dependence, independence and connectivity on proof labels: such relations are proved to be sound, interrelated, and linked to the broader notions of concurrency and causality of events. Additionally, the authors outline the difference between various kinds of bisimulations, such as the history preserving bisimulation for \ccs.

Formal verification has become a cornerstone in the development of new systems, certifying the correctness of their syntax, semantics and behavioral properties in the most reliable way. In the case of concurrent calculi, there is a long tradition which spans from the early mechanizations by \cite{nesi92} and \cite{melham94} in HOL to the recent contributions of the Concurrent Calculi Formalisation Benchmark \cite{ccfb24}; we also recall the Rocq formalization of the $\pi$-calculus by Honsell et al.~\cite{honsell01tcs}, which has been the baseline for numerous higher-order abstract syntax (HOAS) \cite{pfenning88pldi} mechanizations. When it comes to reversible concurrent calculi, however, the landscape looks rather different. Despite the availability of C\# and Java implementations of \ccsk \cite{Cox2009}\cite{aubertb23}, no machine-checked formalization of reversible concurrent calculi currently exists -- at least to the best of our knowledge.

This paper presents the first formalization of \pccsk in Beluga \cite{pientka10}. Our encoding covers the core definitions of the system: its syntax, semantics, and the relations of dependence, independence and connectivity on proof labels. Additionally, this work formalizes the central results presented in Sections 3 and 4 of \cite{aubert25}, including the complementarity of dependence and independence and the relationship between connectivity of transitions and proof labels. The proofs come along with a library of auxiliary lemmas regarding processes, keys and transitions.

Formalizations typically require small adjustments to fit the proof assistant’s framework, while carefully addressing any of the details which are taken for granted in the informal proof. This encoding is no exception: the formalization process led to minor refinements in the definition of connectivity over proof labels and clarified the proof of one of the aforementioned results, which called for a different approach separating base and inductive cases. Beluga, as a proof assistant, is ideal for reasoning about deductive systems together with their meta-theory, as it naturally supports encodings of object-level binding constructs through higher-order abstract syntax and allows pairing terms with the contexts that give them meaning \cite{pientka10}. Although our HOAS encoding leverages Beluga's strengths and showcases its versatility, it also deals with limitations such as the lack of syntactic sugar for existentials or conjunctions; considerations on its adoption are further elaborated in the conclusions.

The paper is structured as follows. Section~\ref{sec:ccskp} provides an informal description of \pccsk and the results under our study. Section~\ref{sec:beluga} presents the Beluga formalization of such notions and properties. Section~\ref{sec:conclusions} contains a technical overview of the formalization, a summary of our contributions and possible future work directions.
The artifact is archived in Zenodo~\cite{cecilia25} and available in the associated GitHub repository \url{https://github.com/CinRC/A-Beluga-Formalization-of-CCSKP}.

%% file: sections/ccskp.tex
\section{\texorpdfstring{\pccsk}{CCSKP}}\label{sec:ccskp}

\newcommand{\spawn}{\sigma}

In this section we recall the main definitions and properties of \pccsk, as outlined in \cite{aubert25}. We assume familiarity with the basic notions of \ccs. Definitions are hyperlinked to their encoding in the repository.

\subsection{Syntax}\label{subsec:ccskp-syn}
As in the standard \ccs, we assume the existence of an infinite set $\names$ of \baselink{1_definitions.bel}{5}{7}{names}, ranged over by $a,b,c$, with a bijection $\overline{\cdot}$: $\names \rightarrow \overline{\names}$
denoting the \emph{complement} of a name; 
names and complementary names respectively denote input and output ports for processes.
We define the set of \baselink{1_definitions.bel}{15}{20}{labels} $\labelset$ as $\names \cup \overline{\names} \cup \{\tau\}$, where $\tau$ denotes the interaction of concurrent processes. $\labelset$ is ranged over by $\alpha$, while $\labelset \setminus \{\tau\}$ is ranged over by $\lambda$.

To introduce reversibility, \ccsk extends the syntax of \ccs with a denumerable set $\keyset$ of \baselink{1_definitions.bel}{9}{13}{keys}, ranged over by $k,m,n$. Labels are paired with keys to define \emph{keyed labels}, which are elements of the cartesian product $\labelset \times \keyset$ and are represented as $a[k],b[m]$; the set of keyed labels is also denoted as $\labelset_{\keyset}$.

\baselink{1_definitions.bel}{22}{30}{Processes} are defined as in the ordinary \ccs, with the addition of keyed prefixes and without operators for recursion or replication:
\begin{center}\label{def:proc}
\begin{tabular}{ l l l l l }
  $X, Y \ \vcentcolon \vcentcolon =$ & \ \ $\textbf{0}$                     & (Inactive)      & $\mid$ $\alpha.X$                 & (Prefix)  \\
                                                           & $\mid$ ${\alpha}[k].X$ & (Keyed prefix)          & $\mid$ $X+Y$ & (Sum)  \\
                                                           & $\mid$ $X \mid Y$ & (Parallel composition)   & $\mid$ $X\setminus a$ & (Restriction)
\end{tabular}
\end{center}

\noindent The set of processes is denoted as $\mathbb{X}$. When preceded by a (keyed) prefix, the inactive process $\textbf{0}$ is usually omitted; the binding power of the operators, from highest to lowest, is $\setminus a,\alpha[k],\alpha,\mid$ and $+$. In restrictions $X\setminus a$,
the occurrences of the name $a$ in $X$ are said to be bound; all other occurrences of names and keys in processes are considered free. Processes that are $\alpha$-equivalent, i.e., that differ only in the choice of their bound names, will be identified. Unlike \cite{aubert25}, restrictions only bind names and not complementary names: this choice does not rule out any significant process (since $X\setminus a$ or $X \setminus \overline{a}$ have the same behaviour) and leads to a clearer correspondence between processes and their encoding.

The set of keys occurring in a process is denoted as $\keys{X}$. A process for which $\keys{X}$ is empty is said to be \baselink{1_definitions.bel}{115}{122}{standard}: in this case, we write that $\std{X}$ holds.

\subsection{Semantics}\label{subsec:ccskp-sem}
The key feature of \pccsk is the notion of \baselink{1_definitions.bel}{65}{73}{proof keyed labels}:
\begin{equation*}
  \theta \ \vcentcolon \vcentcolon = v\alpha[k] \quad \mid \quad v\langle|_{\L}v_1\lambda[k],\,|_{\R}v_2\overline{\lambda}[k]\rangle
\end{equation*}
where
$v, v_1$ and $v_2$ range over strings of symbols $\{|_{\L},|_{\R},+_{\L},+_{\R}\}$.
We denote the set of proof keyed labels as $\kplabelset$, and refer to its elements simply as \enquote{proof labels} for brevity.
The following functions \baselink{1_definitions.bel}{75}{83}{$\ell$} and \baselink{1_definitions.bel}{85}{93}{$\kayop$} map each proof label to its underlying label and key, respectively:
\begin{align*}
	\ell(v \alpha[k]) & = \alpha &   &   & \ell(v \bpair{|_{\L}v_{1}\lambda[k]}{|_{\R}v_{2} \out{\lambda}[k]}) & = \tau &&& 
	\kay{v \alpha[k]} & = k      &   &   & \kay{v \bpair{|_{\L}v_{1}\lambda[k]}{|_{\R}v_{2} \out{\lambda}[k]}} & = k
\end{align*}

\noindent Semantics is given by the \baselink{1_definitions.bel}{135}{173}{labelled transition system} $(\mathbb{X},\kplabelset,\pr{fb}[\ensuremath{\scalebox{0.88}{\theta}}])$, where $\pr{fb}[\ensuremath{\scalebox{0.88}{\theta}}]$ denotes the union of the forward and backward transitions displayed in Fig.~\ref{fig:lts}.
We will refer to the union of forward and backward transitions as \baselink{1_definitions.bel}{175}{179}{combined} transitions.
Given a transition $X \pr{fb}[\ensuremath{\scalebox{0.88}{\theta}}] Y$, the process $X$ is said to be its \emph{source}, while $Y$ is said to be its \emph{target}.

\begin{figure}[ht]
	\begin{tcolorbox}[title = {Prefix and Keyed Prefix}, sidebyside]
		\begin{tcolorbox}[adjusted title=Forward]
			\begin{prooftree}
				\hypo{}
				\infer[left label={\(\std{X}%
					\)}]1[pref]{\alpha. X \pr{f}[\ensuremath{\scalebox{0.88}{\alpha}}][k]  \alpha[k].X}
			\end{prooftree}
			\\[.9em]
			\begin{prooftree}
				\hypo{X \pr{f}[\ensuremath{\scalebox{0.88}{\theta}}] X'}
				\infer[left label={\(\kay{\theta} \neq k\)}]1[kpref]{\alpha[k]. X \pr{f}[\ensuremath{\scalebox{0.88}{\theta}}] \alpha[k].X'}
			\end{prooftree}
		\end{tcolorbox}
		\tcblower
		\begin{tcolorbox}[adjusted title=Backward]
			\begin{prooftree}
				\hypo{}
				\infer[left label={\(\std{X}
					\)}]1[\tRev{pref}]{ \alpha[k].X \pr{b}[\ensuremath{\scalebox{0.88}{\alpha}}][k] \alpha. X}
			\end{prooftree}
			\\[.9em]
			\begin{prooftree}
				\hypo{X' \pr{b}[\ensuremath{\scalebox{0.88}{\theta}}] X}
				\infer[left label={\(\kay{\theta} \neq k\)}]1[\tRev{kpref}]{\alpha[k].X'\pr{b}[\ensuremath{\scalebox{0.88}{\theta}}] \alpha[k]. X }
			\end{prooftree}
		\end{tcolorbox}
	\end{tcolorbox}
	
	\begin{tcolorbox}[adjusted title=Sum, sidebyside]
		\begin{tcolorbox}[adjusted title=Forward]
			\begin{prooftree}
				\hypo{X \pr{f}[\ensuremath{\scalebox{0.88}{\theta}}] X'}
				\infer[left label={\(\std{Y}\)}]1[\( \lplusl \)]{X + Y \pr{f}[\lplusl \ensuremath{\scalebox{0.88}{\theta}}] X' + Y}
			\end{prooftree}
		\end{tcolorbox}
		\tcblower
		\begin{tcolorbox}[adjusted title=Backward]
			\begin{prooftree}
				\hypo{X' \pr{b}[\ensuremath{\scalebox{0.88}{\theta}}] X}
				\infer[left label={\(\std{Y}\)}]1[\(\rlplusl\)]{X' + Y \pr{b}[\lplusl \ensuremath{\scalebox{0.88}{\theta}}] X + Y}
			\end{prooftree}
		\end{tcolorbox}
	\end{tcolorbox}

\begin{tcolorbox}[title = Parallel Composition, sidebyside]
		\begin{tcolorbox}[adjusted title=Forward]
			\begin{prooftree}
				\hypo{X \pr{f}[\ensuremath{\scalebox{0.88}{\theta}}] X'}
				\infer[left label={\(\kay{\theta} \notin \keys{Y}\)}]
				1[\(\lmidl\)]{X \Par Y \pr{f}[\lmidl \ensuremath{\scalebox{0.88}{\theta}}] X' \Par  Y}
			\end{prooftree}
			\\[.9em]
			\begin{prooftree}
				\hypo{X \pr{f}[v_{\L} \ensuremath{\scalebox{0.88}{\lambda}}][k]  X'}
				\hypo{Y \pr{f}[v_{\R} \out{\ensuremath{\scalebox{0.88}{\lambda}}}][k]  Y'}
				\infer2[syn]{X \Par  Y \pr{f}[\cpair{v_{\L} \ensuremath{\scalebox{0.88}{\lambda}} \protect{[k]}}{v_{\R} \out{\ensuremath{\scalebox{0.88}{\lambda}}} \protect{[k]}}] X' \Par  Y'}
			\end{prooftree}
		\end{tcolorbox}
		\tcblower
		\begin{tcolorbox}[adjusted title=Backward]
			\begin{prooftree}
				\hypo{X' \pr{b}[\ensuremath{\scalebox{0.88}{\theta}}] X}
				\infer[left label={\(\kay{\theta} \notin \keys{Y}\)}]
				1[\rlmidl]{X' \Par Y \pr{b}[\lmidl\ensuremath{\scalebox{0.88}{\theta}}] X \Par  Y}
			\end{prooftree}
			\\[.9em]
			\begin{prooftree}
				\hypo{X' \pr{b}[v_{\L} \ensuremath{\scalebox{0.88}{\lambda}}][k]  X}
				\hypo{Y' \pr{b}[v_{\R} \out{\ensuremath{\scalebox{0.88}{\lambda}}}][k]  Y}
				\infer%
				2[\tRev{syn}]{X' \Par  Y' \pr{b}[\cpair{v_{\L} \ensuremath{\scalebox{0.88}{\lambda}} \protect{[k]}}{v_{\R} \out{\ensuremath{\scalebox{0.88}{\lambda}}} \protect{[k]}}] X \Par  Y}
			\end{prooftree}
		\end{tcolorbox}
	\end{tcolorbox}

	\begin{tcolorbox}[adjusted title=Restriction, sidebyside]
		\begin{tcolorbox}[adjusted title=Forward]
			\begin{prooftree}
				\hypo{ X \pr{f}[\ensuremath{\scalebox{0.88}{\theta}}] X '}
				\infer[left label={\(\labl{\theta} \notin \{a, \out{a}\}\)}]1[nu]{X  \bs a  \pr{f}[\ensuremath{\scalebox{0.88}{\theta}}] X ' \bs a}
			\end{prooftree}
		\end{tcolorbox}
		\tcblower
		\begin{tcolorbox}[adjusted title=Backward]
			\begin{prooftree}
				\hypo{ X' \pr{b}[\ensuremath{\scalebox{0.88}{\theta}}] X}
				\infer[left label={\(\labl{\theta} \notin \{a, \out{a}\}\)}]1[\tRev{nu}]{X ' \bs a\pr{b}[\ensuremath{\scalebox{0.88}{\theta}}] X  \bs a  }
			\end{prooftree}
		\end{tcolorbox}
	\end{tcolorbox}
\vspace{-4mm}
	\caption{Forward and backward transition rules for \pccsk (right rules for $\mid$ and $+$ omitted).}
	\label{fig:lts}
\end{figure}

\vspace{-4mm}
\begin{example}\label{ex:one}

\emph{Consider a webpage that allows the user to interact via two independent buttons: one to toggle between light and dark mode, and another to switch between two different languages. The initial state of the system can be modeled as the parallel composition $m \mid l$, where the labels $m$ and $l$ represent the actions to switch the visual mode and the language, respectively.}

\emph{The action of changing the visual mode can be represented by the following forward transition: $\,m \mid l \ \pr{f}[|_L m][k] \ m[k] \mid l$. The target process preserves the label $m$ and pairs it with a fresh key $k$. The proof label $\,|_L m[k]\,$ not only stores the label $m$ and key $k$ used in the transition, but also indicates that the action occurred on the left-hand side of a parallel composition.}

\emph{Suppose the user now wishes to revert to the previous visual mode: pressing the mode button again can be interpreted as undoing the previously executed action. This can be modeled by the following backward transition, which removes the key associated with the earlier forward step: $\,m[k] \mid l \ \pr{b}[|_L m][k] \ m \mid l$.}

\end{example}

Two transitions are said to be \emph{composable} if they can be performed consecutively -- that is, the target of the first transition is the source of the second transition.  A \baselink{1_definitions.bel}{181}{186}{path} is a (potentially empty) sequence of composable transitions and can be denoted as $X \pr{fb}[]^* Y$, where $X$ is the source of the first transition (also called the \emph{source} of the path) and $Y$ is the target of the last transition (also called the \emph{target} of the path); in other words, $\pr{fb}^*$ is the reflexive and transitive closure of $\pr{fb}$. A process $X$ is \baselink{1_definitions.bel}{188}{191}{reachable} if there exists a path whose target is $X$ and whose source is a standard process. This process, which can be proved to be unique (cf.\ Lemma B.13 in \cite{aubert24}), is called the \emph{origin} of $X$ and is denoted as $O_X$.

Reachability allows to rule out faulty processes which are syntactically well-formed, but whose particular selection of keys is inconsistent. For instance, this arises when the same key denotes successive actions, as in the process $a[k].b[k]$, or when keys internally form a cycle, as in the deadlocked process $a[k].b[m] \mid \bar{b}[m].\bar{a}[k]$, where neither action can be undone because of the presence of its associated key in the other thread. From this point on, each process will be assumed to be reachable.

The \baselink{2_basic_properties.bel}{160}{192}{loop lemma} (cf.\ Lemma 3.8 in \cite{aubert25}) is an important result characterizing reversible labelled transition systems. It states that any transition $X \pr{fb}[\ensuremath{\scalebox{0.88}{\theta}}] Y$ can be reversed, yielding a transition $Y \pr{fb}[\ensuremath{\scalebox{0.88}{\theta}}] X$; moreover, the reversing operator is an involution (i.e., reversing a transition twice returns the original transition). The validity of the loop lemma follows directly from the symmetry of the LTS (Labelled Transition System) rules presented in Fig.~\ref{fig:lts}.

\begin{figure}[H]
	\begin{tcolorbox}[title = {Connectivity Relation}, fontupper=\linespread{.9}\small, sidebyside, sidebyside align=top]
		\begin{tcolorbox}[adjusted title=Action]
			\raisebox{0em}{
				\makebox[.4\textwidth][c]{
					\begin{prooftree}
						\hypo{}
						\infer[]1[A\(^1\)]{\alpha[k] \conn \theta}
					\end{prooftree}
				}
			}
			\raisebox{0em}{
				\makebox[.4\textwidth][c]{
					\begin{prooftree}
						\hypo{}
						\infer[]1[A\(^2\)]{\theta \conn \alpha[k]}
					\end{prooftree}
				}
			}
		\end{tcolorbox}
		\begin{tcolorbox}[adjusted title=Parallel]
			\makebox[.4\textwidth][c]{
				\begin{prooftree}
					\hypo{\theta_1 \conn \theta_2}
					\infer[]1[P\(^1_{\Dir}\)]{\lmidd \theta_1 \conn \lmidd \theta_2}
				\end{prooftree}
			}
			\makebox[.4\textwidth][c]{
				\begin{prooftree}
					\hypo{ }
					\infer[]1[P\(^2_{\Dir}\)]{\lmidd \theta_1 \conn \lmidod \theta_2}
				\end{prooftree}
			}
		\end{tcolorbox}
		\tcblower
		\begin{tcolorbox}[adjusted title=Choice]
			\makebox[.4\textwidth][c]{
				\begin{prooftree}
					\hypo{\theta_1 \conn \theta_2}
					\infer[]1[C\(^1_{\Dir}\)]{\lplusd \theta_1 \conn \lplusd \theta_2}
				\end{prooftree}
			}
			\makebox[.4\textwidth][c]{
				\begin{prooftree}
					\hypo{}
					\infer[]1[C\(^2_{\Dir}\)]{\lplusd \theta_1 \conn \lplusod \theta_2}
				\end{prooftree}
			}
		\end{tcolorbox}
		
		\begin{tcolorbox}[adjusted title=Synchronization]
			\begin{prooftree}
				\hypo{\theta \conn \theta_{\D}}
				\infer[]1[S\(^1_{\Dir}\)]{\lmidd \theta  \conn \cpair{\theta_{\L}}{\theta_{\R}}}
			\end{prooftree}
			\hfill
			\begin{prooftree}
				\hypo{\theta_{\D} \conn \theta}
				\infer[]1[S\(^2_{\Dir}\)]{\cpair{\theta_{\L}}{\theta_{\R}} \conn \lmidd \theta}
			\end{prooftree}
			\\[.8em]
			\begin{prooftree}
				\hypo{\theta_1 \conn \theta'_1}
				\hypo{\theta_2 \conn \theta'_2}
				\infer[]2[S\(^3\)]{\cpair {\theta_1} {\theta_2} \conn \cpair {\theta'_1} {\theta'_2}}
			\end{prooftree}
		\end{tcolorbox}
	\end{tcolorbox}
	\begin{multicols}{2}
		\begin{tcolorbox}[title = {Dependence Relation}, fontupper=\linespread{.9}\small]
			\begin{tcolorbox}[adjusted title=Action]
				\raisebox{-1.25em}{ 
					\makebox[.4\textwidth][c]{
						\begin{prooftree}
							\hypo{}
							\infer[]1[A\(^1\)]{\alpha[k] \sdep \theta}
						\end{prooftree}
					}
				}
				\raisebox{-1.25em}{ 
					\makebox[.4\textwidth][c]{
						\begin{prooftree}
							\hypo{}
							\infer[]1[A\(^2\)]{\theta \sdep \alpha[k]}
						\end{prooftree}
					}
				}
			\end{tcolorbox}
			\begin{tcolorbox}[adjusted title=Choice]
				\makebox[.4\textwidth][c]{
					\begin{prooftree}
						\hypo{\theta \sdep \theta'}
						\infer[]1[C\(^1_{\Dir}\)]{\lplusd \theta \sdep \lplusd \theta'}
					\end{prooftree}
				}
				\makebox[.4\textwidth][c]{
					\begin{prooftree}
						\hypo{}
						\infer[]1[C\(^2_{\Dir}\)]{\lplusd \theta \sdep \lplusod \theta'}
					\end{prooftree}
				}
			\end{tcolorbox}
			\begin{tcolorbox}[adjusted title=Parallel]
				\makebox[.4\textwidth][c]{
					\begin{prooftree}
						\hypo{\theta \sdep \theta'}
						\infer[]1[P\(^1_{\Dir}\)]{\lmidd \theta \sdep\ \lmidd \theta'}
					\end{prooftree}
				}
				\makebox[.4\textwidth][c]{
					\begin{prooftree}
						\hypo{\kay{\theta} = \kay{\theta'}}
						\infer[]1[P\(^2_{\Dir}\)]{\lmidd \theta \sdep\ \lmidod \theta'}
					\end{prooftree}
				}
			\end{tcolorbox}
			\begin{tcolorbox}[adjusted title=Synchronization]
				\begin{prooftree}
					\hypo{\theta \sdep \theta_{\D}}
					\infer[]1[S\(^1_{\Dir}\)]{\lmidd \theta  \sdep \cpair {\theta_{\L}} {\theta_{\R}}}
				\end{prooftree}
				\hfill
				\begin{prooftree}
					\hypo{\theta_{\D} \sdep \theta}
					\infer[]1[S\(^2_{\Dir}\)]{\cpair {\theta_{\L}} {\theta_{\R}} \sdep\ \lmidd \theta}
				\end{prooftree}
				\\[.8em]
				\begin{prooftree}
					\hypo{\theta_i \sdep \theta'_i}
					\hypo{\theta_j \conn \theta'_j}
					\hypo{i, j \in \{1, 2\}, i \neq j}
					\infer[]3[S\(^3\)]{\cpair {\theta_1} {\theta_2} \sdep \cpair {\theta'_1} {\theta'_2}}
				\end{prooftree}
			\end{tcolorbox}
		\end{tcolorbox}
		\begin{tcolorbox}[title = {Independence Relation}, fontupper=\linespread{.9}\small]
			\begin{tcolorbox}[adjusted title = {Action}]
				\vbox to 23
				pt {\vfil
					\hfill \emph{(empty)} \hfill~
					\vfil
				}
			\end{tcolorbox}
			\begin{tcolorbox}[adjusted title=Choice]
				\makebox[.4\textwidth][c]{
					\begin{prooftree}
						\hypo{\theta \ind  \theta'}
						\infer[]1[C\(^1_{\Dir}\)]{\lplusd \theta \ind  \lplusd \theta'}
					\end{prooftree}
				}
				\makebox[.4\textwidth][c]{ 
				}
			\end{tcolorbox}
			\begin{tcolorbox}[adjusted title=Parallel]
				\makebox[.4\textwidth][c]{
					\begin{prooftree}
						\hypo{\theta \ind  \theta'}
						\infer[]1[P\(^1_{\Dir}\)]{\lmidd \theta \ind\  \lmidd \theta'}
					\end{prooftree}
				}
				\makebox[.4\textwidth][c]{
					\begin{prooftree}
						\hypo{\kay{\theta} \neq \kay{\theta'}}
						\infer[]1[P\(^2_{\Dir}\)]{\lmidd \theta \ind \ \lmidod \theta'}
					\end{prooftree}
				}
			\end{tcolorbox}
			\begin{tcolorbox}[adjusted title=Synchronization]
				\begin{prooftree}
					\hypo{\theta \ind  \theta_{\D}}
					\infer[]1[S\(^1_{\Dir}\)]{\lmidd \theta  \ind  \cpair{\theta_{\L}}{\theta_{\R}}}
				\end{prooftree}
				\hfill
				\begin{prooftree}
					\hypo{\theta_{\D} \ind \theta}
					\infer[]1[S\(^2_{\Dir}\)]{\cpair{\theta_{\L}}{\theta_{\R}} \ind\ \lmidd \theta}
				\end{prooftree}
				\\[.8em]
				\begin{prooftree}
					\hypo{\theta_1 \ind  \theta'_1}
					\hypo{\theta_2 \ind  \theta'_2}
					\infer[]2[S\(^3\)]{\cpair {\theta_1} {\theta_2} \ind  \cpair {\theta'_1} {\theta'_2}}
				\end{prooftree}
			\end{tcolorbox}
		\end{tcolorbox}
	\end{multicols}
	\vspace{-5mm}
	\caption{Causality relations on proof labels.
	}
	\label{fig:pccsk-relations}
\end{figure}

Finally, the binary relations of \baselink{1_definitions.bel}{201}{218}{connectivity}, \baselink{1_definitions.bel}{220}{238}{dependence} and \baselink{1_definitions.bel}{240}{253}{independence} on proof labels, respectively denoted as $\conn,\sdep$ and $\ind$, are defined by the rules displayed in Fig.~\ref{fig:pccsk-relations}, where the label $\Dir$ ranges over $\{\L,\R\}$ and $\overline{\Dir}$ denotes the opposite of $\Dir$ (i.e., $\overline{\L} = \R$ and $\overline{\R} = \L$). Such relations will be referred to as \emph{causality relations} for brevity. Compared to \cite{aubert25}, the rule A$^2$ for connectivity and dependence has been slightly modified, ensuring that each relation is symmetric and simplifying their encoding. This comes at the cost of losing uniqueness in derivations of judgements such as $\theta_1 \conn \theta_2$; however, this property has been shown not to be required for the purposes of our development.

\begin{example}\label{ex:two}

\emph{The process $m \mid l$, introduced in Example~\ref{ex:one} to model a webpage, can perform a forward transition labelled by $|_L m[k_1]$, representing the toggling of the visual mode. It can also perform a transition $\,m \mid l \ \pr{f}[|_R l][k_2] \ m \mid l[k_2]$, denoting the change of the language of the webpage. The two transitions are independent, as the order in which they are executed does not affect the resulting state. This is reflected in the independence of the two proof labels $|_L m[k_1]$ and $|_R l[k_2]$, which follows from the P$^2_{\L}$ rule in Fig.~\ref{fig:pccsk-relations}.}

\emph{Conversely, consider the process $a.b \ | \ \bar{b}$. It can perform a forward transition labelled by $|_L a[k]$, followed by another forward transition labelled by $|_L b[n]$. However, these transitions cannot be performed in reverse order, since the input action along $b$ is only enabled after the input action along $a$ has occurred; the two transitions are thus causally related. This is reflected in the dependence of the two proof labels $|_L a[k]$ and $|_L b[n]$, which follows from the P$^1_{\L}$ and A$^1$ rules in Fig.~\ref{fig:pccsk-relations}.}
\end{example}

\subsection{Properties of causality relations}\label{subsec:rel-prop}

We now turn to the theorems and lemmas object of our study. Their complete proof can be found in \cite{aubert24}, the technical report accompanying \cite{aubert25}.
The following theorem specifies the relationship between connectivity of transitions and connectivity of proof labels:

\begin{theorem}[cf.\ Proposition 4.4 in \cite{aubert25}]
\label{thm:conn} \mbox{}\\
\vspace{-4mm}
\begin{enumerate}[label=(\roman*)]
	\item If \(t_1 : X_1 \pr{fb}[\ensuremath{\scalebox{0.88}{\theta}}_1] X'_1\) and \(t_2 : X_2 \pr{fb}[\ensuremath{\scalebox{0.88}{\theta}}_2] X'_2\) are connected, then \(\theta_1 \conn \theta_2\). \extlink{4_connectivity_relationship_one.bel}{1}{291} \label{thm:conn-one}
	\item If \(\theta_1 \conn \theta_2\), then there exist \(t_1 : X_1 \pr{fb}[\ensuremath{\scalebox{0.88}{\theta}}_1] X'_1\) and \(t_2 : X_2 \pr{fb}[\ensuremath{\scalebox{0.88}{\theta}}_2] X'_2\) such that \(t_1\) and \(t_2\) are connected. \extlink{6_connectivity_relationship_two.bel}{590}{598} \label{thm:conn-two}
\end{enumerate}
\end{theorem}

\noindent The proof of Theorem~\theoremitemref{thm:conn}{thm:conn-one} relies on the fact that $O_{X_1}=O_{X_2}$ and proceeds by induction over such origin process: recall that each process is assumed to be reachable and, therefore, has an origin. The equality of $O_{X_1}$ and $O_{X_2}$ follows from the two lemmas:

\begin{lemma}
	\label{lem:path-origin}
	For all reachable processes \(X\) and \(Y\), there exists a path \(X \pr{fb}^*Y \) iff \(\orig{X} = \orig{Y}\).
\end{lemma}

\begin{lemma}
	\label{lem:connected-origin}
	If \(t_1 : X_1 \pr{fb}[\ensuremath{\scalebox{0.88}{\theta}}_1] X'_1\) and \(t_2 : X_2 \pr{fb}[\ensuremath{\scalebox{0.88}{\theta}}_2] X'_2\) are connected,  then \(\orig{X_1} = \orig{X_2}\).
\end{lemma}

\noindent Conversely, the proof of Theorem~\theoremitemref{thm:conn}{thm:conn-two} proceeds by structural induction over the given hypothesis \(\theta_1 \conn \theta_2\) and relies on the following:

\begin{definition}[Realisation]\label{def:real}
	A process \(X\) \emph{realises the proof label \(\theta\)} if there exist \(X_1\) and \(X_2\) such that \(X \pr{fb}^* X_1 \pr{fb}[\ensuremath{\scalebox{0.88}{\theta}}] X_2\). \extlink{5_lemmas_connectivity_relationship_two.bel}{3}{8}
\end{definition}

\begin{lemma}
	\label{lem:realis}
	For every proof label \(\theta\), there exists a process that realises it, and we denote it \(\real{\theta}\). \extlink{5_lemmas_connectivity_relationship_two.bel}{10}{29}
\end{lemma}

\noindent Next, the following theorem states the complementarity of the dependence and independence relations:

\begin{theorem}[cf.\ Theorem 4.9 in \cite{aubert25}]
\label{thm:compl} \mbox{}\\
For all \(\theta_1\), \(\theta_2\),
\begin{enumerate}[label=(\roman*)]
	\item If $\theta_1 \ind \theta_2$ then $\theta_1 \conn \theta_2$. \extlink{7_complementarity.bel}{81}{98} \label{thm:compl-one}
	\item If $\theta_1 \sdep \theta_2$ then $\theta_1 \conn \theta_2$. \extlink{7_complementarity.bel}{100}{121} \label{thm:compl-two}
	\item If $\theta_1 \conn \theta_2$ then either $\theta_1 \ind \theta_2$ or $\theta_1 \sdep \theta_2$, but not both. \extlink{7_complementarity.bel}{131}{212}\label{thm:compl-three}
\end{enumerate}
\end{theorem}

\noindent This theorem is proved by induction over the structure of the given binary relation.

%% file: sections/beluga.tex
\section{Beluga Formalization}\label{sec:beluga}

In this section we outline the key points of the Beluga formalization of the notions presented in Section~\ref{sec:ccskp}. Definitions and proofs omitted for brevity are hyperlinked to their encoding in the repository.

\subsection{Syntax}
Beluga is structured in two layers: the LF (Logical Frameworks \cite{harper93jacm}) level, which is used to specify the formal system under study, and the computation level, which supports programming with LF data \cite{pientka10}. To encode the syntax of our system, only the former level is deployed.
Names, keys, labels and processes are encoded using the LF types
displayed in Fig.~\ref{fig:basetypes}.
\begin{figure}[ht]
\begin{lstlisting}[style=belugastyleframes,linerange={BaseTypes-End}]
%%% CCSK^P %%%

%%% Syntax %%%

%% BaseTypes %%
LF names: type =;                LF proc: type =
LF keys: type =                    | null: proc                              % 0
  | z: keys                        | pref: labels → proc → proc            % A.X
  | s: keys → keys;               | kpref: labels → keys → proc → proc   % A[k].X
LF labels: type =                  | sum: proc → proc → proc               % X+Y
  | inp: names → labels           | par: proc → proc → proc               % X|Y
  | out: names → labels           | nu: (names → proc) → proc;            % X\a
  | tau: labels;
%% End %%

%% Context %%
schema ctx = names;
%% End %%
% Equality of processes
LF eqp: proc → proc → type =
  | refp: eqp X X
;

% Equality of labels
LF eql: labels → labels → type =
  | refl: eql A A
;

% Equality of keys
LF eqk: keys → keys → type =
  | refk: eqk K K
;

% Strict order relation on keys (nats)
LF less: keys → keys → type =
  | lzero: {K:keys} less z (s K)
  | lsucc: less K M → less (s K) (s M)
;

% Inequality of keys
LF neq: keys → keys → type =
  | nless: less K M → neq K M
  | ngreat: less M K → neq K M
;

%%% Semantics %%%

%% ProofKeyedLabels %%
LF pr_lab: type =                         LF lab: pr_lab → labels → type =
  | pr_base: labels → keys → pr_lab       | lab_base: lab (pr_base A K) A
  | pr_suml: pr_lab → pr_lab               | lab_suml: lab T A → lab (pr_suml T) A
  | pr_sumr: pr_lab → pr_lab               | lab_sumr: lab T A → lab (pr_sumr T) A
  | pr_parl: pr_lab → pr_lab               | lab_parl: lab T A → lab (pr_parl T) A
  | pr_parr: pr_lab → pr_lab               | lab_parr: lab T A → lab (pr_parr T) A
  | pr_sync: pr_lab → pr_lab → pr_lab;    | lab_sync: lab (pr_sync T1 T2) tau;
LF valid: pr_lab → type =
  | v_base: valid (pr_base A K)
  | v_suml: valid T → valid (pr_suml T)
  | v_sumr: valid T → valid (pr_sumr T)
  | v_parl: valid T → valid (pr_parl T)
  | v_parr: valid T → valid (pr_parr T)
  | v_synl: valid T1 → valid T2 → lab T1 (inp A) → key T1 K
         → lab T2 (out A) → key T2 K → valid (pr_sync T1 T2)
  | v_synr: valid T1 → valid T2 → lab T1 (out A) → key T1 K
         → lab T2 (inp A) → key T2 K → valid (pr_sync T1 T2);
%% End %%

LF key: pr_lab → keys → type =
  | key_base: key (pr_base A K) K
  | key_suml: key T K → key (pr_suml T) K
  | key_sumr: key T K → key (pr_sumr T) K
  | key_parl: key T K → key (pr_parl T) K
  | key_parr: key T K → key (pr_parr T) K
  | key_sync: key T1 K → key (pr_sync T1 T2) K
;

% Standard processes
LF std: proc → type =
  | std_null: std null
  | std_pref: std X → std (pref A X)
  | std_sum: std X → std Y → std (sum X Y)
  | std_par: std X → std Y → std (par X Y)
  | std_nu: ({a:names} std (X a)) → std (nu X)
;

% Non occurrence of keys in processes
LF notin: keys → proc → type =
  | not_null: notin K null
  | not_pref: notin K X → notin K (pref A X)
  | not_kpref: neq K M → notin K X → notin K (kpref A M X)
  | not_sum: notin K X → notin K Y → notin K (sum X Y)
  | not_par: notin K X → notin K Y → notin K (par X Y)
  | not_nu: ({a:names} notin K (X a)) → notin K (nu X)
;

%% Forward %%
LF fstep: proc → pr_lab → proc → type =
  | fs_pref: std X → fstep (pref A X) (pr_base A K) (kpref A K X)
  | fs_kpref: fstep X T X' → key T M → neq K M
           → fstep (kpref A K X) T (kpref A K X')
  | fs_suml: fstep X T X' → std Y → fstep (sum X Y) (pr_suml T) (sum X' Y)
  | fs_sumr: fstep Y T Y' → std X → fstep (sum X Y) (pr_sumr T) (sum X Y')
  | fs_parl: fstep X T X' → key T K → notin K Y
          → fstep (par X Y) (pr_parl T) (par X' Y)
  | fs_parr: fstep Y T Y' → key T K → notin K X
          → fstep (par X Y) (pr_parr T) (par X Y')
  | fs_synl: fstep X T1 X' → lab T1 (inp L) → key T1 K
          → fstep Y T2 Y' → lab T2 (out L) → key T2 K
          → fstep (par X Y) (pr_sync T1 T2) (par X' Y')
  | fs_synr: fstep X T1 X' → lab T1 (out L) → key T1 K
          → fstep Y T2 Y' → lab T2 (inp L) → key T2 K
          → fstep (par X Y) (pr_sync T1 T2) (par X' Y')
  | fs_nu: ({a:names} fstep (X a) T (X' a)) → fstep (nu X) T (nu X');
%% End %%

% Backward transitions
LF bstep: proc → pr_lab → proc → type =
  | bs_pref: std X → bstep (kpref A K X) (pr_base A K) (pref A X)
  | bs_kpref: bstep X' T X → key T M → neq K M → bstep (kpref A K X') T (kpref A K X)
  | bs_suml: bstep X' T X → std Y → bstep (sum X' Y) (pr_suml T) (sum X Y)
  | bs_sumr: bstep Y' T Y → std X → bstep (sum X Y') (pr_sumr T) (sum X Y)
  | bs_parl: bstep X' T X → key T K → notin K Y → bstep (par X' Y) (pr_parl T) (par X Y)
  | bs_parr: bstep Y' T Y → key T K → notin K X → bstep (par X Y') (pr_parr T) (par X Y)
  | bs_synl: bstep X' T1 X → lab T1 (inp L) → key T1 K → bstep Y' T2 Y → lab T2 (out L) → key T2 K → bstep (par X' Y') (pr_sync T1 T2) (par X Y)
  | bs_synr: bstep X' T1 X → lab T1 (out L) → key T1 K → bstep Y' T2 Y → lab T2 (inp L) → key T2 K → bstep (par X' Y') (pr_sync T1 T2) (par X Y)
  | bs_nu: ({a:names} bstep (X' a) T (X a)) → bstep (nu X') T (nu X)
;

%% Combined %%
LF step: proc → pr_lab → proc → type =
  | fw: fstep X T X' → step X T X'
  | bw: bstep X' T X → step X' T X;
%% End %%

%% Paths %%
LF step*: proc → proc → type =
  | id_s*: step* X X
  | st_s*: step X T Y → step* X Y
  | tr_s*: step* X Y → step* Y Z → step* X Z;
LF reachable: proc → type =
  | rch: std X → step* X Y → reachable Y;
LF conn_tr: step X T1 X' → step Y T2 Y' → type =
  | ct: {S1:step X T1 X'}{S2: step Y T2 Y'} step* X Y' → conn_tr S1 S2;
%% End %%

% In the definition of pr_lab, we admit "spurious" proof labels <T1,T2> for any T1 and T2:
% however, according to the informal definition of proof keyed labels, <T1,T2> is defined 
% only for those T1 and T2 which have the same key and complementary labels.
% As a consequence, in this encoding, there exist proof keyed labels which are not labeling
% any transition: results like realization of each proof label do not hold anymore.
% We define a new type family characterizing only the "valid" proof keyed labels.

% Type family to encode valid proof labels

%%% Connectivity, dependence and independence of proof keyed labels %%%

%% Connectivity %%
LF conn: pr_lab → pr_lab → type =
  | c_a1: conn (pr_base A K) T
  | c_a2: conn T (pr_base A K)
  | c_c1l: conn T1 T2 → conn (pr_suml T1) (pr_suml T2)
  | c_c1r: conn T1 T2 → conn (pr_sumr T1) (pr_sumr T2)
  | c_c2l: conn (pr_suml T1) (pr_sumr T2)
  | c_c2r: conn (pr_sumr T1) (pr_suml T2)
  | c_p1l: conn T1 T2 → conn (pr_parl T1) (pr_parl T2)
  | c_p1r: conn T1 T2 → conn (pr_parr T1) (pr_parr T2)
  | c_p2l: conn (pr_parl T1) (pr_parr T2)
  | c_p2r: conn (pr_parr T1) (pr_parl T2)
  | c_s1l: conn T TL → conn (pr_parl T) (pr_sync TL TR)
  | c_s1r: conn T TR → conn (pr_parr T) (pr_sync TL TR)
  | c_s2l: conn TL T → conn (pr_sync TL TR) (pr_parl T)
  | c_s2r: conn TR T → conn (pr_sync TL TR) (pr_parr T)
  | c_s3: conn T1 T1' → conn T2 T2' → conn (pr_sync T1 T2) (pr_sync T1' T2');
%% End %%

% Dependence
LF dep: pr_lab → pr_lab → type =
  | d_a1: dep (pr_base A K) T
  | d_a2: dep T (pr_base A K)
  | d_c1l: dep T1 T2 → dep (pr_suml T1) (pr_suml T2)
  | d_c1r: dep T1 T2 → dep (pr_sumr T1) (pr_sumr T2)
  | d_c2l: dep (pr_suml T1) (pr_sumr T2)
  | d_c2r: dep (pr_sumr T1) (pr_suml T2)
  | d_p1l: dep T1 T2 → dep (pr_parl T1) (pr_parl T2)
  | d_p1r: dep T1 T2 → dep (pr_parr T1) (pr_parr T2)
  | d_p2l: key T1 K → key T2 K → dep (pr_parl T1) (pr_parr T2)
  | d_p2r: key T1 K → key T2 K → dep (pr_parr T1) (pr_parl T2)
  | d_s1l: dep T TL → dep (pr_parl T) (pr_sync TL TR)
  | d_s1r: dep T TR → dep (pr_parr T) (pr_sync TL TR)
  | d_s2l: dep TL T → dep (pr_sync TL TR) (pr_parl T)
  | d_s2r: dep TR T → dep (pr_sync TL TR) (pr_parr T)
  | d_s3l: dep T1 T1' → conn T2 T2' → dep (pr_sync T1 T2) (pr_sync T1' T2')
  | d_s3r: conn T1 T1' → dep T2 T2' → dep (pr_sync T1 T2) (pr_sync T1' T2')
;

% Independence
LF indep: pr_lab → pr_lab → type =
  | i_c1l: indep T1 T2 → indep (pr_suml T1) (pr_suml T2)
  | i_c1r: indep T1 T2 → indep (pr_sumr T1) (pr_sumr T2)
  | i_p1l: indep T1 T2 → indep (pr_parl T1) (pr_parl T2)
  | i_p1r: indep T1 T2 → indep (pr_parr T1) (pr_parr T2)
  | i_p2l: key T1 K → key T2 M → neq K M → indep (pr_parl T1) (pr_parr T2)
  | i_p2r: key T1 K → key T2 M → neq K M → indep (pr_parr T1) (pr_parl T2)
  | i_s1l: indep T TL → indep (pr_parl T) (pr_sync TL TR)
  | i_s1r: indep T TR → indep (pr_parr T) (pr_sync TL TR)
  | i_s2l: indep TL T → indep (pr_sync TL TR) (pr_parl T)
  | i_s2r: indep TR T → indep (pr_sync TL TR) (pr_parr T)
  | i_s3: indep T1 T1' → indep T2 T2' → indep (pr_sync T1 T2) (pr_sync T1' T2')
;
\end{lstlisting}
\vspace{-0.5\baselineskip}
\caption{Encoding of the syntax of \pccsk.}
\label{fig:basetypes}
\end{figure}

Since names in \pccsk are an infinite set without any additional assumption, they are represented by a type \type{names} without constructors;
as explained in \cite{momigliano24}, this type will be dynamically inhabited by variables introduced through contexts. This is enabled by the following line of code:
\begin{lstlisting}[linerange={Context-End}]
%%% CCSK^P %%%

%%% Syntax %%%

%% BaseTypes %%
LF names: type =;                LF proc: type =
LF keys: type =                    | null: proc                              % 0
  | z: keys                        | pref: labels → proc → proc            % A.X
  | s: keys → keys;               | kpref: labels → keys → proc → proc   % A[k].X
LF labels: type =                  | sum: proc → proc → proc               % X+Y
  | inp: names → labels           | par: proc → proc → proc               % X|Y
  | out: names → labels           | nu: (names → proc) → proc;            % X\a
  | tau: labels;
%% End %%

%% Context %%
schema ctx = names;
%% End %%
% Equality of processes
LF eqp: proc → proc → type =
  | refp: eqp X X
;

% Equality of labels
LF eql: labels → labels → type =
  | refl: eql A A
;

% Equality of keys
LF eqk: keys → keys → type =
  | refk: eqk K K
;

% Strict order relation on keys (nats)
LF less: keys → keys → type =
  | lzero: {K:keys} less z (s K)
  | lsucc: less K M → less (s K) (s M)
;

% Inequality of keys
LF neq: keys → keys → type =
  | nless: less K M → neq K M
  | ngreat: less M K → neq K M
;

%%% Semantics %%%

%% ProofKeyedLabels %%
LF pr_lab: type =                         LF lab: pr_lab → labels → type =
  | pr_base: labels → keys → pr_lab       | lab_base: lab (pr_base A K) A
  | pr_suml: pr_lab → pr_lab               | lab_suml: lab T A → lab (pr_suml T) A
  | pr_sumr: pr_lab → pr_lab               | lab_sumr: lab T A → lab (pr_sumr T) A
  | pr_parl: pr_lab → pr_lab               | lab_parl: lab T A → lab (pr_parl T) A
  | pr_parr: pr_lab → pr_lab               | lab_parr: lab T A → lab (pr_parr T) A
  | pr_sync: pr_lab → pr_lab → pr_lab;    | lab_sync: lab (pr_sync T1 T2) tau;
LF valid: pr_lab → type =
  | v_base: valid (pr_base A K)
  | v_suml: valid T → valid (pr_suml T)
  | v_sumr: valid T → valid (pr_sumr T)
  | v_parl: valid T → valid (pr_parl T)
  | v_parr: valid T → valid (pr_parr T)
  | v_synl: valid T1 → valid T2 → lab T1 (inp A) → key T1 K
         → lab T2 (out A) → key T2 K → valid (pr_sync T1 T2)
  | v_synr: valid T1 → valid T2 → lab T1 (out A) → key T1 K
         → lab T2 (inp A) → key T2 K → valid (pr_sync T1 T2);
%% End %%

LF key: pr_lab → keys → type =
  | key_base: key (pr_base A K) K
  | key_suml: key T K → key (pr_suml T) K
  | key_sumr: key T K → key (pr_sumr T) K
  | key_parl: key T K → key (pr_parl T) K
  | key_parr: key T K → key (pr_parr T) K
  | key_sync: key T1 K → key (pr_sync T1 T2) K
;

% Standard processes
LF std: proc → type =
  | std_null: std null
  | std_pref: std X → std (pref A X)
  | std_sum: std X → std Y → std (sum X Y)
  | std_par: std X → std Y → std (par X Y)
  | std_nu: ({a:names} std (X a)) → std (nu X)
;

% Non occurrence of keys in processes
LF notin: keys → proc → type =
  | not_null: notin K null
  | not_pref: notin K X → notin K (pref A X)
  | not_kpref: neq K M → notin K X → notin K (kpref A M X)
  | not_sum: notin K X → notin K Y → notin K (sum X Y)
  | not_par: notin K X → notin K Y → notin K (par X Y)
  | not_nu: ({a:names} notin K (X a)) → notin K (nu X)
;

%% Forward %%
LF fstep: proc → pr_lab → proc → type =
  | fs_pref: std X → fstep (pref A X) (pr_base A K) (kpref A K X)
  | fs_kpref: fstep X T X' → key T M → neq K M
           → fstep (kpref A K X) T (kpref A K X')
  | fs_suml: fstep X T X' → std Y → fstep (sum X Y) (pr_suml T) (sum X' Y)
  | fs_sumr: fstep Y T Y' → std X → fstep (sum X Y) (pr_sumr T) (sum X Y')
  | fs_parl: fstep X T X' → key T K → notin K Y
          → fstep (par X Y) (pr_parl T) (par X' Y)
  | fs_parr: fstep Y T Y' → key T K → notin K X
          → fstep (par X Y) (pr_parr T) (par X Y')
  | fs_synl: fstep X T1 X' → lab T1 (inp L) → key T1 K
          → fstep Y T2 Y' → lab T2 (out L) → key T2 K
          → fstep (par X Y) (pr_sync T1 T2) (par X' Y')
  | fs_synr: fstep X T1 X' → lab T1 (out L) → key T1 K
          → fstep Y T2 Y' → lab T2 (inp L) → key T2 K
          → fstep (par X Y) (pr_sync T1 T2) (par X' Y')
  | fs_nu: ({a:names} fstep (X a) T (X' a)) → fstep (nu X) T (nu X');
%% End %%

% Backward transitions
LF bstep: proc → pr_lab → proc → type =
  | bs_pref: std X → bstep (kpref A K X) (pr_base A K) (pref A X)
  | bs_kpref: bstep X' T X → key T M → neq K M → bstep (kpref A K X') T (kpref A K X)
  | bs_suml: bstep X' T X → std Y → bstep (sum X' Y) (pr_suml T) (sum X Y)
  | bs_sumr: bstep Y' T Y → std X → bstep (sum X Y') (pr_sumr T) (sum X Y)
  | bs_parl: bstep X' T X → key T K → notin K Y → bstep (par X' Y) (pr_parl T) (par X Y)
  | bs_parr: bstep Y' T Y → key T K → notin K X → bstep (par X Y') (pr_parr T) (par X Y)
  | bs_synl: bstep X' T1 X → lab T1 (inp L) → key T1 K → bstep Y' T2 Y → lab T2 (out L) → key T2 K → bstep (par X' Y') (pr_sync T1 T2) (par X Y)
  | bs_synr: bstep X' T1 X → lab T1 (out L) → key T1 K → bstep Y' T2 Y → lab T2 (inp L) → key T2 K → bstep (par X' Y') (pr_sync T1 T2) (par X Y)
  | bs_nu: ({a:names} bstep (X' a) T (X a)) → bstep (nu X') T (nu X)
;

%% Combined %%
LF step: proc → pr_lab → proc → type =
  | fw: fstep X T X' → step X T X'
  | bw: bstep X' T X → step X' T X;
%% End %%

%% Paths %%
LF step*: proc → proc → type =
  | id_s*: step* X X
  | st_s*: step X T Y → step* X Y
  | tr_s*: step* X Y → step* Y Z → step* X Z;
LF reachable: proc → type =
  | rch: std X → step* X Y → reachable Y;
LF conn_tr: step X T1 X' → step Y T2 Y' → type =
  | ct: {S1:step X T1 X'}{S2: step Y T2 Y'} step* X Y' → conn_tr S1 S2;
%% End %%

% In the definition of pr_lab, we admit "spurious" proof labels <T1,T2> for any T1 and T2:
% however, according to the informal definition of proof keyed labels, <T1,T2> is defined 
% only for those T1 and T2 which have the same key and complementary labels.
% As a consequence, in this encoding, there exist proof keyed labels which are not labeling
% any transition: results like realization of each proof label do not hold anymore.
% We define a new type family characterizing only the "valid" proof keyed labels.

% Type family to encode valid proof labels

%%% Connectivity, dependence and independence of proof keyed labels %%%

%% Connectivity %%
LF conn: pr_lab → pr_lab → type =
  | c_a1: conn (pr_base A K) T
  | c_a2: conn T (pr_base A K)
  | c_c1l: conn T1 T2 → conn (pr_suml T1) (pr_suml T2)
  | c_c1r: conn T1 T2 → conn (pr_sumr T1) (pr_sumr T2)
  | c_c2l: conn (pr_suml T1) (pr_sumr T2)
  | c_c2r: conn (pr_sumr T1) (pr_suml T2)
  | c_p1l: conn T1 T2 → conn (pr_parl T1) (pr_parl T2)
  | c_p1r: conn T1 T2 → conn (pr_parr T1) (pr_parr T2)
  | c_p2l: conn (pr_parl T1) (pr_parr T2)
  | c_p2r: conn (pr_parr T1) (pr_parl T2)
  | c_s1l: conn T TL → conn (pr_parl T) (pr_sync TL TR)
  | c_s1r: conn T TR → conn (pr_parr T) (pr_sync TL TR)
  | c_s2l: conn TL T → conn (pr_sync TL TR) (pr_parl T)
  | c_s2r: conn TR T → conn (pr_sync TL TR) (pr_parr T)
  | c_s3: conn T1 T1' → conn T2 T2' → conn (pr_sync T1 T2) (pr_sync T1' T2');
%% End %%

% Dependence
LF dep: pr_lab → pr_lab → type =
  | d_a1: dep (pr_base A K) T
  | d_a2: dep T (pr_base A K)
  | d_c1l: dep T1 T2 → dep (pr_suml T1) (pr_suml T2)
  | d_c1r: dep T1 T2 → dep (pr_sumr T1) (pr_sumr T2)
  | d_c2l: dep (pr_suml T1) (pr_sumr T2)
  | d_c2r: dep (pr_sumr T1) (pr_suml T2)
  | d_p1l: dep T1 T2 → dep (pr_parl T1) (pr_parl T2)
  | d_p1r: dep T1 T2 → dep (pr_parr T1) (pr_parr T2)
  | d_p2l: key T1 K → key T2 K → dep (pr_parl T1) (pr_parr T2)
  | d_p2r: key T1 K → key T2 K → dep (pr_parr T1) (pr_parl T2)
  | d_s1l: dep T TL → dep (pr_parl T) (pr_sync TL TR)
  | d_s1r: dep T TR → dep (pr_parr T) (pr_sync TL TR)
  | d_s2l: dep TL T → dep (pr_sync TL TR) (pr_parl T)
  | d_s2r: dep TR T → dep (pr_sync TL TR) (pr_parr T)
  | d_s3l: dep T1 T1' → conn T2 T2' → dep (pr_sync T1 T2) (pr_sync T1' T2')
  | d_s3r: conn T1 T1' → dep T2 T2' → dep (pr_sync T1 T2) (pr_sync T1' T2')
;

% Independence
LF indep: pr_lab → pr_lab → type =
  | i_c1l: indep T1 T2 → indep (pr_suml T1) (pr_suml T2)
  | i_c1r: indep T1 T2 → indep (pr_sumr T1) (pr_sumr T2)
  | i_p1l: indep T1 T2 → indep (pr_parl T1) (pr_parl T2)
  | i_p1r: indep T1 T2 → indep (pr_parr T1) (pr_parr T2)
  | i_p2l: key T1 K → key T2 M → neq K M → indep (pr_parl T1) (pr_parr T2)
  | i_p2r: key T1 K → key T2 M → neq K M → indep (pr_parr T1) (pr_parl T2)
  | i_s1l: indep T TL → indep (pr_parl T) (pr_sync TL TR)
  | i_s1r: indep T TR → indep (pr_parr T) (pr_sync TL TR)
  | i_s2l: indep TL T → indep (pr_sync TL TR) (pr_parl T)
  | i_s2r: indep TR T → indep (pr_sync TL TR) (pr_parr T)
  | i_s3: indep T1 T1' → indep T2 T2' → indep (pr_sync T1 T2) (pr_sync T1' T2')
;
\end{lstlisting}
This line declares contexts made of a finite collection of distinct variables of type \type{names}, identified via the keyword \type{ctx}. Thanks to this setup, we can work with \emph{contextual processes} of the form \verb|[g |$\vdash$\verb| X]|, i.e., processes \verb|X| whose free names are drawn from the context \verb|g|. Contextual objects live in the computation level.

Keys are by assumption denumerable, and the LTS rules for keyed prefixes require equality of keys to be decidable. Both conditions are satisfied by encoding keys explicitly as natural numbers.\footnote{Note that properties such as decidability of equality must be stated and proved manually, as Beluga does not provide a built-in library of properties of natural numbers.} An alternative approach would be to rely on contexts, as is done for names: however, this would require managing mixed contexts of names and keys, and having a more complex encoding of transitions and paths.

Restrictions $X\setminus a$ are represented by terms of the form \verb|(nu \a.(X a))|, where \verb|\x.(f x)| is Beluga's notation for functions \verb|f| mapping \verb|x| to \verb|f(x)|: following the higher-order abstract syntax (HOAS) paradigm, the bound name $a$ is represented as the implicit argument of a meta-language function \verb|\a.(X a)| from \type{names} to \type{proc}. In this way, we leverage the meta-language implementation of binders to achieve $\alpha$-renaming and capture-avoiding substitutions for free.

An important but often overlooked aspect of formalizations is the \emph{adequacy} of the encoding: the encoding must constitute a faithful representation of the original system into study~\cite{cheneynv12}. Adequacy is generally established by proving the existence of a compositional bijection between the mathematical model and its formalized counterpart. The discussion of the adequacy of our encoding is omitted due to space constraints.

\subsection{Semantics}\label{sec:beluga-semantics}

Proof labels are encoded by the type \type{pr_lab} in Fig.~\ref{fig:sem}. Rather than directly modeling the informal definition of proof labels, by defining strings over the symbols $\{|_{\L},|_{\R},+_{\L},+_{\R}\}$ as lists, we are introducing four constructors (\verb|pr_suml|, \verb|pr_sumr|, etc.) that build proof labels incrementally by appending one symbol at a time. This provides a stronger induction principle and streamlines the encoding of LTS rules and subsequent proofs.

\begin{figure}[htbp]
\begin{lstlisting}[style=belugastyleframes,linerange={ProofKeyedLabels-End,Forward-End}]
%%% CCSK^P %%%

%%% Syntax %%%

%% BaseTypes %%
LF names: type =;                LF proc: type =
LF keys: type =                    | null: proc                              % 0
  | z: keys                        | pref: labels → proc → proc            % A.X
  | s: keys → keys;               | kpref: labels → keys → proc → proc   % A[k].X
LF labels: type =                  | sum: proc → proc → proc               % X+Y
  | inp: names → labels           | par: proc → proc → proc               % X|Y
  | out: names → labels           | nu: (names → proc) → proc;            % X\a
  | tau: labels;
%% End %%

%% Context %%
schema ctx = names;
%% End %%
% Equality of processes
LF eqp: proc → proc → type =
  | refp: eqp X X
;

% Equality of labels
LF eql: labels → labels → type =
  | refl: eql A A
;

% Equality of keys
LF eqk: keys → keys → type =
  | refk: eqk K K
;

% Strict order relation on keys (nats)
LF less: keys → keys → type =
  | lzero: {K:keys} less z (s K)
  | lsucc: less K M → less (s K) (s M)
;

% Inequality of keys
LF neq: keys → keys → type =
  | nless: less K M → neq K M
  | ngreat: less M K → neq K M
;

%%% Semantics %%%

%% ProofKeyedLabels %%
LF pr_lab: type =                         LF lab: pr_lab → labels → type =
  | pr_base: labels → keys → pr_lab       | lab_base: lab (pr_base A K) A
  | pr_suml: pr_lab → pr_lab               | lab_suml: lab T A → lab (pr_suml T) A
  | pr_sumr: pr_lab → pr_lab               | lab_sumr: lab T A → lab (pr_sumr T) A
  | pr_parl: pr_lab → pr_lab               | lab_parl: lab T A → lab (pr_parl T) A
  | pr_parr: pr_lab → pr_lab               | lab_parr: lab T A → lab (pr_parr T) A
  | pr_sync: pr_lab → pr_lab → pr_lab;    | lab_sync: lab (pr_sync T1 T2) tau;
LF valid: pr_lab → type =
  | v_base: valid (pr_base A K)
  | v_suml: valid T → valid (pr_suml T)
  | v_sumr: valid T → valid (pr_sumr T)
  | v_parl: valid T → valid (pr_parl T)
  | v_parr: valid T → valid (pr_parr T)
  | v_synl: valid T1 → valid T2 → lab T1 (inp A) → key T1 K
         → lab T2 (out A) → key T2 K → valid (pr_sync T1 T2)
  | v_synr: valid T1 → valid T2 → lab T1 (out A) → key T1 K
         → lab T2 (inp A) → key T2 K → valid (pr_sync T1 T2);
%% End %%

LF key: pr_lab → keys → type =
  | key_base: key (pr_base A K) K
  | key_suml: key T K → key (pr_suml T) K
  | key_sumr: key T K → key (pr_sumr T) K
  | key_parl: key T K → key (pr_parl T) K
  | key_parr: key T K → key (pr_parr T) K
  | key_sync: key T1 K → key (pr_sync T1 T2) K
;

% Standard processes
LF std: proc → type =
  | std_null: std null
  | std_pref: std X → std (pref A X)
  | std_sum: std X → std Y → std (sum X Y)
  | std_par: std X → std Y → std (par X Y)
  | std_nu: ({a:names} std (X a)) → std (nu X)
;

% Non occurrence of keys in processes
LF notin: keys → proc → type =
  | not_null: notin K null
  | not_pref: notin K X → notin K (pref A X)
  | not_kpref: neq K M → notin K X → notin K (kpref A M X)
  | not_sum: notin K X → notin K Y → notin K (sum X Y)
  | not_par: notin K X → notin K Y → notin K (par X Y)
  | not_nu: ({a:names} notin K (X a)) → notin K (nu X)
;

%% Forward %%
LF fstep: proc → pr_lab → proc → type =
  | fs_pref: std X → fstep (pref A X) (pr_base A K) (kpref A K X)
  | fs_kpref: fstep X T X' → key T M → neq K M
           → fstep (kpref A K X) T (kpref A K X')
  | fs_suml: fstep X T X' → std Y → fstep (sum X Y) (pr_suml T) (sum X' Y)
  | fs_sumr: fstep Y T Y' → std X → fstep (sum X Y) (pr_sumr T) (sum X Y')
  | fs_parl: fstep X T X' → key T K → notin K Y
          → fstep (par X Y) (pr_parl T) (par X' Y)
  | fs_parr: fstep Y T Y' → key T K → notin K X
          → fstep (par X Y) (pr_parr T) (par X Y')
  | fs_synl: fstep X T1 X' → lab T1 (inp L) → key T1 K
          → fstep Y T2 Y' → lab T2 (out L) → key T2 K
          → fstep (par X Y) (pr_sync T1 T2) (par X' Y')
  | fs_synr: fstep X T1 X' → lab T1 (out L) → key T1 K
          → fstep Y T2 Y' → lab T2 (inp L) → key T2 K
          → fstep (par X Y) (pr_sync T1 T2) (par X' Y')
  | fs_nu: ({a:names} fstep (X a) T (X' a)) → fstep (nu X) T (nu X');
%% End %%

% Backward transitions
LF bstep: proc → pr_lab → proc → type =
  | bs_pref: std X → bstep (kpref A K X) (pr_base A K) (pref A X)
  | bs_kpref: bstep X' T X → key T M → neq K M → bstep (kpref A K X') T (kpref A K X)
  | bs_suml: bstep X' T X → std Y → bstep (sum X' Y) (pr_suml T) (sum X Y)
  | bs_sumr: bstep Y' T Y → std X → bstep (sum X Y') (pr_sumr T) (sum X Y)
  | bs_parl: bstep X' T X → key T K → notin K Y → bstep (par X' Y) (pr_parl T) (par X Y)
  | bs_parr: bstep Y' T Y → key T K → notin K X → bstep (par X Y') (pr_parr T) (par X Y)
  | bs_synl: bstep X' T1 X → lab T1 (inp L) → key T1 K → bstep Y' T2 Y → lab T2 (out L) → key T2 K → bstep (par X' Y') (pr_sync T1 T2) (par X Y)
  | bs_synr: bstep X' T1 X → lab T1 (out L) → key T1 K → bstep Y' T2 Y → lab T2 (inp L) → key T2 K → bstep (par X' Y') (pr_sync T1 T2) (par X Y)
  | bs_nu: ({a:names} bstep (X' a) T (X a)) → bstep (nu X') T (nu X)
;

%% Combined %%
LF step: proc → pr_lab → proc → type =
  | fw: fstep X T X' → step X T X'
  | bw: bstep X' T X → step X' T X;
%% End %%

%% Paths %%
LF step*: proc → proc → type =
  | id_s*: step* X X
  | st_s*: step X T Y → step* X Y
  | tr_s*: step* X Y → step* Y Z → step* X Z;
LF reachable: proc → type =
  | rch: std X → step* X Y → reachable Y;
LF conn_tr: step X T1 X' → step Y T2 Y' → type =
  | ct: {S1:step X T1 X'}{S2: step Y T2 Y'} step* X Y' → conn_tr S1 S2;
%% End %%

% In the definition of pr_lab, we admit "spurious" proof labels <T1,T2> for any T1 and T2:
% however, according to the informal definition of proof keyed labels, <T1,T2> is defined 
% only for those T1 and T2 which have the same key and complementary labels.
% As a consequence, in this encoding, there exist proof keyed labels which are not labeling
% any transition: results like realization of each proof label do not hold anymore.
% We define a new type family characterizing only the "valid" proof keyed labels.

% Type family to encode valid proof labels

%%% Connectivity, dependence and independence of proof keyed labels %%%

%% Connectivity %%
LF conn: pr_lab → pr_lab → type =
  | c_a1: conn (pr_base A K) T
  | c_a2: conn T (pr_base A K)
  | c_c1l: conn T1 T2 → conn (pr_suml T1) (pr_suml T2)
  | c_c1r: conn T1 T2 → conn (pr_sumr T1) (pr_sumr T2)
  | c_c2l: conn (pr_suml T1) (pr_sumr T2)
  | c_c2r: conn (pr_sumr T1) (pr_suml T2)
  | c_p1l: conn T1 T2 → conn (pr_parl T1) (pr_parl T2)
  | c_p1r: conn T1 T2 → conn (pr_parr T1) (pr_parr T2)
  | c_p2l: conn (pr_parl T1) (pr_parr T2)
  | c_p2r: conn (pr_parr T1) (pr_parl T2)
  | c_s1l: conn T TL → conn (pr_parl T) (pr_sync TL TR)
  | c_s1r: conn T TR → conn (pr_parr T) (pr_sync TL TR)
  | c_s2l: conn TL T → conn (pr_sync TL TR) (pr_parl T)
  | c_s2r: conn TR T → conn (pr_sync TL TR) (pr_parr T)
  | c_s3: conn T1 T1' → conn T2 T2' → conn (pr_sync T1 T2) (pr_sync T1' T2');
%% End %%

% Dependence
LF dep: pr_lab → pr_lab → type =
  | d_a1: dep (pr_base A K) T
  | d_a2: dep T (pr_base A K)
  | d_c1l: dep T1 T2 → dep (pr_suml T1) (pr_suml T2)
  | d_c1r: dep T1 T2 → dep (pr_sumr T1) (pr_sumr T2)
  | d_c2l: dep (pr_suml T1) (pr_sumr T2)
  | d_c2r: dep (pr_sumr T1) (pr_suml T2)
  | d_p1l: dep T1 T2 → dep (pr_parl T1) (pr_parl T2)
  | d_p1r: dep T1 T2 → dep (pr_parr T1) (pr_parr T2)
  | d_p2l: key T1 K → key T2 K → dep (pr_parl T1) (pr_parr T2)
  | d_p2r: key T1 K → key T2 K → dep (pr_parr T1) (pr_parl T2)
  | d_s1l: dep T TL → dep (pr_parl T) (pr_sync TL TR)
  | d_s1r: dep T TR → dep (pr_parr T) (pr_sync TL TR)
  | d_s2l: dep TL T → dep (pr_sync TL TR) (pr_parl T)
  | d_s2r: dep TR T → dep (pr_sync TL TR) (pr_parr T)
  | d_s3l: dep T1 T1' → conn T2 T2' → dep (pr_sync T1 T2) (pr_sync T1' T2')
  | d_s3r: conn T1 T1' → dep T2 T2' → dep (pr_sync T1 T2) (pr_sync T1' T2')
;

% Independence
LF indep: pr_lab → pr_lab → type =
  | i_c1l: indep T1 T2 → indep (pr_suml T1) (pr_suml T2)
  | i_c1r: indep T1 T2 → indep (pr_sumr T1) (pr_sumr T2)
  | i_p1l: indep T1 T2 → indep (pr_parl T1) (pr_parl T2)
  | i_p1r: indep T1 T2 → indep (pr_parr T1) (pr_parr T2)
  | i_p2l: key T1 K → key T2 M → neq K M → indep (pr_parl T1) (pr_parr T2)
  | i_p2r: key T1 K → key T2 M → neq K M → indep (pr_parr T1) (pr_parl T2)
  | i_s1l: indep T TL → indep (pr_parl T) (pr_sync TL TR)
  | i_s1r: indep T TR → indep (pr_parr T) (pr_sync TL TR)
  | i_s2l: indep TL T → indep (pr_sync TL TR) (pr_parl T)
  | i_s2r: indep TR T → indep (pr_sync TL TR) (pr_parr T)
  | i_s3: indep T1 T1' → indep T2 T2' → indep (pr_sync T1 T2) (pr_sync T1' T2')
;
\end{lstlisting}
\vspace{-0.5\baselineskip}
\caption{Encoding of the semantics of \pccsk.}
\label{fig:sem}
\end{figure}

\vspace{-0.5\baselineskip}
In Beluga, predicates are encoded as type families, i.e., types parametrized by arguments: a predicate $P(x_1, \ldots, x_n)$ holds iff the corresponding type (\texttt{P x}$_{\texttt{1}}$ \ldots~\texttt{x}$_{\texttt{n}}$) is not empty. Type families are also used to encode functions, identified with their graph, as in the case of the functions $\ell$ and $\kayop$ returning the label and key of a proof label: the former is encoded by the type family \type{lab} in Fig.~\ref{fig:sem}, while the latter is encoded by the type family \linktype{1_definitions.bel}{85}{93}{key},
here omitted for brevity. For example, given a proof label $\theta$ and a label $\alpha$, represented as \verb|T| and \verb|A| in the encoding, the type {\color{belugagreen} \verb|lab T A|} is inhabited iff $\ell(\theta)=\alpha$.

Our encoding of proof labels pays the price of being over-expressive: the constructor \verb|pr_sync| accepts any two proof labels regardless of their key or label, generating terms that fall outside the original definition. For example, the term \texttt{(pr\_sync (pr\_base A K) (pr\_base B M))} has type \type{pr\_lab} for any labels \texttt{A}, \texttt{B} and keys \texttt{K}, \texttt{M}, while its counterpart $\langle|_{\L} \alpha[k],\,|_{\R}\beta[m]\rangle$ is well-defined only if $\beta = \overline{\alpha}$ and $k=m$. While this is harmless in most of our development, since such spurious terms do not label any actual transition, it becomes an issue when proving theorems universally quantified on proof labels, such as Lemma~\ref{lem:realis}. To address this problem, we introduce an additional predicate \type{valid}, displayed in Fig.~\ref{fig:sem}, which filters out the spurious terms.
It can be proved the existence of a bijection between proof labels and the set of elements \verb|T| of type {\type{pr_lab} for which \verb|valid T| holds. From this point forward, we will refer to such terms as \emph{valid} proof labels.

Forward and backward LTS rules are defined through the type families \type{fstep} and \linktype{1_definitions.bel}{155}{173}{bstep} in Fig.~\ref{fig:sem} (with the latter omitted here for brevity). These rules rely on the additional type families \linktype{1_definitions.bel}{115}{122}{std}, \linktype{1_definitions.bel}{124}{132}{notin} and \linktype{1_definitions.bel}{50}{60}{neq}, hyperlinked to their formalization in the repository, which respectively hold when a process is standard, when a key does not occur in a process, and when two keys are not equal. The parameters \verb|X| and \verb|X'| in the \verb|fs_nu| rule are functions from \type{names} to \type{proc}, whose arguments represent the binders of the restrictions. The universal quantification \verb|{a:names}| is used to abstract over the particular choice of the binder; moreover, $a$ or $\bar{a}$ does not occur in the proof label \verb|T|, since such parameter does not depend on \verb|a| within the body of the universal quantification.

Combined transitions, paths, reachable processes and connected transitions are defined as follows:

\begin{lstlisting}[linerange={Combined-End,Paths-End}]
%%% CCSK^P %%%

%%% Syntax %%%

%% BaseTypes %%
LF names: type =;                LF proc: type =
LF keys: type =                    | null: proc                              % 0
  | z: keys                        | pref: labels → proc → proc            % A.X
  | s: keys → keys;               | kpref: labels → keys → proc → proc   % A[k].X
LF labels: type =                  | sum: proc → proc → proc               % X+Y
  | inp: names → labels           | par: proc → proc → proc               % X|Y
  | out: names → labels           | nu: (names → proc) → proc;            % X\a
  | tau: labels;
%% End %%

%% Context %%
schema ctx = names;
%% End %%
% Equality of processes
LF eqp: proc → proc → type =
  | refp: eqp X X
;

% Equality of labels
LF eql: labels → labels → type =
  | refl: eql A A
;

% Equality of keys
LF eqk: keys → keys → type =
  | refk: eqk K K
;

% Strict order relation on keys (nats)
LF less: keys → keys → type =
  | lzero: {K:keys} less z (s K)
  | lsucc: less K M → less (s K) (s M)
;

% Inequality of keys
LF neq: keys → keys → type =
  | nless: less K M → neq K M
  | ngreat: less M K → neq K M
;

%%% Semantics %%%

%% ProofKeyedLabels %%
LF pr_lab: type =                         LF lab: pr_lab → labels → type =
  | pr_base: labels → keys → pr_lab       | lab_base: lab (pr_base A K) A
  | pr_suml: pr_lab → pr_lab               | lab_suml: lab T A → lab (pr_suml T) A
  | pr_sumr: pr_lab → pr_lab               | lab_sumr: lab T A → lab (pr_sumr T) A
  | pr_parl: pr_lab → pr_lab               | lab_parl: lab T A → lab (pr_parl T) A
  | pr_parr: pr_lab → pr_lab               | lab_parr: lab T A → lab (pr_parr T) A
  | pr_sync: pr_lab → pr_lab → pr_lab;    | lab_sync: lab (pr_sync T1 T2) tau;
LF valid: pr_lab → type =
  | v_base: valid (pr_base A K)
  | v_suml: valid T → valid (pr_suml T)
  | v_sumr: valid T → valid (pr_sumr T)
  | v_parl: valid T → valid (pr_parl T)
  | v_parr: valid T → valid (pr_parr T)
  | v_synl: valid T1 → valid T2 → lab T1 (inp A) → key T1 K
         → lab T2 (out A) → key T2 K → valid (pr_sync T1 T2)
  | v_synr: valid T1 → valid T2 → lab T1 (out A) → key T1 K
         → lab T2 (inp A) → key T2 K → valid (pr_sync T1 T2);
%% End %%

LF key: pr_lab → keys → type =
  | key_base: key (pr_base A K) K
  | key_suml: key T K → key (pr_suml T) K
  | key_sumr: key T K → key (pr_sumr T) K
  | key_parl: key T K → key (pr_parl T) K
  | key_parr: key T K → key (pr_parr T) K
  | key_sync: key T1 K → key (pr_sync T1 T2) K
;

% Standard processes
LF std: proc → type =
  | std_null: std null
  | std_pref: std X → std (pref A X)
  | std_sum: std X → std Y → std (sum X Y)
  | std_par: std X → std Y → std (par X Y)
  | std_nu: ({a:names} std (X a)) → std (nu X)
;

% Non occurrence of keys in processes
LF notin: keys → proc → type =
  | not_null: notin K null
  | not_pref: notin K X → notin K (pref A X)
  | not_kpref: neq K M → notin K X → notin K (kpref A M X)
  | not_sum: notin K X → notin K Y → notin K (sum X Y)
  | not_par: notin K X → notin K Y → notin K (par X Y)
  | not_nu: ({a:names} notin K (X a)) → notin K (nu X)
;

%% Forward %%
LF fstep: proc → pr_lab → proc → type =
  | fs_pref: std X → fstep (pref A X) (pr_base A K) (kpref A K X)
  | fs_kpref: fstep X T X' → key T M → neq K M
           → fstep (kpref A K X) T (kpref A K X')
  | fs_suml: fstep X T X' → std Y → fstep (sum X Y) (pr_suml T) (sum X' Y)
  | fs_sumr: fstep Y T Y' → std X → fstep (sum X Y) (pr_sumr T) (sum X Y')
  | fs_parl: fstep X T X' → key T K → notin K Y
          → fstep (par X Y) (pr_parl T) (par X' Y)
  | fs_parr: fstep Y T Y' → key T K → notin K X
          → fstep (par X Y) (pr_parr T) (par X Y')
  | fs_synl: fstep X T1 X' → lab T1 (inp L) → key T1 K
          → fstep Y T2 Y' → lab T2 (out L) → key T2 K
          → fstep (par X Y) (pr_sync T1 T2) (par X' Y')
  | fs_synr: fstep X T1 X' → lab T1 (out L) → key T1 K
          → fstep Y T2 Y' → lab T2 (inp L) → key T2 K
          → fstep (par X Y) (pr_sync T1 T2) (par X' Y')
  | fs_nu: ({a:names} fstep (X a) T (X' a)) → fstep (nu X) T (nu X');
%% End %%

% Backward transitions
LF bstep: proc → pr_lab → proc → type =
  | bs_pref: std X → bstep (kpref A K X) (pr_base A K) (pref A X)
  | bs_kpref: bstep X' T X → key T M → neq K M → bstep (kpref A K X') T (kpref A K X)
  | bs_suml: bstep X' T X → std Y → bstep (sum X' Y) (pr_suml T) (sum X Y)
  | bs_sumr: bstep Y' T Y → std X → bstep (sum X Y') (pr_sumr T) (sum X Y)
  | bs_parl: bstep X' T X → key T K → notin K Y → bstep (par X' Y) (pr_parl T) (par X Y)
  | bs_parr: bstep Y' T Y → key T K → notin K X → bstep (par X Y') (pr_parr T) (par X Y)
  | bs_synl: bstep X' T1 X → lab T1 (inp L) → key T1 K → bstep Y' T2 Y → lab T2 (out L) → key T2 K → bstep (par X' Y') (pr_sync T1 T2) (par X Y)
  | bs_synr: bstep X' T1 X → lab T1 (out L) → key T1 K → bstep Y' T2 Y → lab T2 (inp L) → key T2 K → bstep (par X' Y') (pr_sync T1 T2) (par X Y)
  | bs_nu: ({a:names} bstep (X' a) T (X a)) → bstep (nu X') T (nu X)
;

%% Combined %%
LF step: proc → pr_lab → proc → type =
  | fw: fstep X T X' → step X T X'
  | bw: bstep X' T X → step X' T X;
%% End %%

%% Paths %%
LF step*: proc → proc → type =
  | id_s*: step* X X
  | st_s*: step X T Y → step* X Y
  | tr_s*: step* X Y → step* Y Z → step* X Z;
LF reachable: proc → type =
  | rch: std X → step* X Y → reachable Y;
LF conn_tr: step X T1 X' → step Y T2 Y' → type =
  | ct: {S1:step X T1 X'}{S2: step Y T2 Y'} step* X Y' → conn_tr S1 S2;
%% End %%

% In the definition of pr_lab, we admit "spurious" proof labels <T1,T2> for any T1 and T2:
% however, according to the informal definition of proof keyed labels, <T1,T2> is defined 
% only for those T1 and T2 which have the same key and complementary labels.
% As a consequence, in this encoding, there exist proof keyed labels which are not labeling
% any transition: results like realization of each proof label do not hold anymore.
% We define a new type family characterizing only the "valid" proof keyed labels.

% Type family to encode valid proof labels

%%% Connectivity, dependence and independence of proof keyed labels %%%

%% Connectivity %%
LF conn: pr_lab → pr_lab → type =
  | c_a1: conn (pr_base A K) T
  | c_a2: conn T (pr_base A K)
  | c_c1l: conn T1 T2 → conn (pr_suml T1) (pr_suml T2)
  | c_c1r: conn T1 T2 → conn (pr_sumr T1) (pr_sumr T2)
  | c_c2l: conn (pr_suml T1) (pr_sumr T2)
  | c_c2r: conn (pr_sumr T1) (pr_suml T2)
  | c_p1l: conn T1 T2 → conn (pr_parl T1) (pr_parl T2)
  | c_p1r: conn T1 T2 → conn (pr_parr T1) (pr_parr T2)
  | c_p2l: conn (pr_parl T1) (pr_parr T2)
  | c_p2r: conn (pr_parr T1) (pr_parl T2)
  | c_s1l: conn T TL → conn (pr_parl T) (pr_sync TL TR)
  | c_s1r: conn T TR → conn (pr_parr T) (pr_sync TL TR)
  | c_s2l: conn TL T → conn (pr_sync TL TR) (pr_parl T)
  | c_s2r: conn TR T → conn (pr_sync TL TR) (pr_parr T)
  | c_s3: conn T1 T1' → conn T2 T2' → conn (pr_sync T1 T2) (pr_sync T1' T2');
%% End %%

% Dependence
LF dep: pr_lab → pr_lab → type =
  | d_a1: dep (pr_base A K) T
  | d_a2: dep T (pr_base A K)
  | d_c1l: dep T1 T2 → dep (pr_suml T1) (pr_suml T2)
  | d_c1r: dep T1 T2 → dep (pr_sumr T1) (pr_sumr T2)
  | d_c2l: dep (pr_suml T1) (pr_sumr T2)
  | d_c2r: dep (pr_sumr T1) (pr_suml T2)
  | d_p1l: dep T1 T2 → dep (pr_parl T1) (pr_parl T2)
  | d_p1r: dep T1 T2 → dep (pr_parr T1) (pr_parr T2)
  | d_p2l: key T1 K → key T2 K → dep (pr_parl T1) (pr_parr T2)
  | d_p2r: key T1 K → key T2 K → dep (pr_parr T1) (pr_parl T2)
  | d_s1l: dep T TL → dep (pr_parl T) (pr_sync TL TR)
  | d_s1r: dep T TR → dep (pr_parr T) (pr_sync TL TR)
  | d_s2l: dep TL T → dep (pr_sync TL TR) (pr_parl T)
  | d_s2r: dep TR T → dep (pr_sync TL TR) (pr_parr T)
  | d_s3l: dep T1 T1' → conn T2 T2' → dep (pr_sync T1 T2) (pr_sync T1' T2')
  | d_s3r: conn T1 T1' → dep T2 T2' → dep (pr_sync T1 T2) (pr_sync T1' T2')
;

% Independence
LF indep: pr_lab → pr_lab → type =
  | i_c1l: indep T1 T2 → indep (pr_suml T1) (pr_suml T2)
  | i_c1r: indep T1 T2 → indep (pr_sumr T1) (pr_sumr T2)
  | i_p1l: indep T1 T2 → indep (pr_parl T1) (pr_parl T2)
  | i_p1r: indep T1 T2 → indep (pr_parr T1) (pr_parr T2)
  | i_p2l: key T1 K → key T2 M → neq K M → indep (pr_parl T1) (pr_parr T2)
  | i_p2r: key T1 K → key T2 M → neq K M → indep (pr_parr T1) (pr_parl T2)
  | i_s1l: indep T TL → indep (pr_parl T) (pr_sync TL TR)
  | i_s1r: indep T TR → indep (pr_parr T) (pr_sync TL TR)
  | i_s2l: indep TL T → indep (pr_sync TL TR) (pr_parl T)
  | i_s2r: indep TR T → indep (pr_sync TL TR) (pr_parr T)
  | i_s3: indep T1 T1' → indep T2 T2' → indep (pr_sync T1 T2) (pr_sync T1' T2')
;
\end{lstlisting}

Paths, or multi-step transitions, can be encoded equivalently using only two constructors; in this development, the more verbose version has been adopted as it simplified the proof search.

Finally, the relations of connectivity, dependence and independence are encoded through three type families \type{conn}, \linktype{1_definitions.bel}{220}{238}{dep} and \linktype{1_definitions.bel}{240}{253}{indep}. The former is displayed in Fig.~\ref{fig:conn}. While some of the rules in Fig.~\ref{fig:pccsk-relations} are grouped together, by using the label $\Dir$ in the place of $\L$ and $\R$, the encoding requires each rule to be stated separately, with its own constructor.

\begin{figure}[htbp]
\begin{lstlisting}[style=belugastyleframes,linerange={Connectivity-End}]
%%% CCSK^P %%%

%%% Syntax %%%

%% BaseTypes %%
LF names: type =;                LF proc: type =
LF keys: type =                    | null: proc                              % 0
  | z: keys                        | pref: labels → proc → proc            % A.X
  | s: keys → keys;               | kpref: labels → keys → proc → proc   % A[k].X
LF labels: type =                  | sum: proc → proc → proc               % X+Y
  | inp: names → labels           | par: proc → proc → proc               % X|Y
  | out: names → labels           | nu: (names → proc) → proc;            % X\a
  | tau: labels;
%% End %%

%% Context %%
schema ctx = names;
%% End %%
% Equality of processes
LF eqp: proc → proc → type =
  | refp: eqp X X
;

% Equality of labels
LF eql: labels → labels → type =
  | refl: eql A A
;

% Equality of keys
LF eqk: keys → keys → type =
  | refk: eqk K K
;

% Strict order relation on keys (nats)
LF less: keys → keys → type =
  | lzero: {K:keys} less z (s K)
  | lsucc: less K M → less (s K) (s M)
;

% Inequality of keys
LF neq: keys → keys → type =
  | nless: less K M → neq K M
  | ngreat: less M K → neq K M
;

%%% Semantics %%%

%% ProofKeyedLabels %%
LF pr_lab: type =                         LF lab: pr_lab → labels → type =
  | pr_base: labels → keys → pr_lab       | lab_base: lab (pr_base A K) A
  | pr_suml: pr_lab → pr_lab               | lab_suml: lab T A → lab (pr_suml T) A
  | pr_sumr: pr_lab → pr_lab               | lab_sumr: lab T A → lab (pr_sumr T) A
  | pr_parl: pr_lab → pr_lab               | lab_parl: lab T A → lab (pr_parl T) A
  | pr_parr: pr_lab → pr_lab               | lab_parr: lab T A → lab (pr_parr T) A
  | pr_sync: pr_lab → pr_lab → pr_lab;    | lab_sync: lab (pr_sync T1 T2) tau;
LF valid: pr_lab → type =
  | v_base: valid (pr_base A K)
  | v_suml: valid T → valid (pr_suml T)
  | v_sumr: valid T → valid (pr_sumr T)
  | v_parl: valid T → valid (pr_parl T)
  | v_parr: valid T → valid (pr_parr T)
  | v_synl: valid T1 → valid T2 → lab T1 (inp A) → key T1 K
         → lab T2 (out A) → key T2 K → valid (pr_sync T1 T2)
  | v_synr: valid T1 → valid T2 → lab T1 (out A) → key T1 K
         → lab T2 (inp A) → key T2 K → valid (pr_sync T1 T2);
%% End %%

LF key: pr_lab → keys → type =
  | key_base: key (pr_base A K) K
  | key_suml: key T K → key (pr_suml T) K
  | key_sumr: key T K → key (pr_sumr T) K
  | key_parl: key T K → key (pr_parl T) K
  | key_parr: key T K → key (pr_parr T) K
  | key_sync: key T1 K → key (pr_sync T1 T2) K
;

% Standard processes
LF std: proc → type =
  | std_null: std null
  | std_pref: std X → std (pref A X)
  | std_sum: std X → std Y → std (sum X Y)
  | std_par: std X → std Y → std (par X Y)
  | std_nu: ({a:names} std (X a)) → std (nu X)
;

% Non occurrence of keys in processes
LF notin: keys → proc → type =
  | not_null: notin K null
  | not_pref: notin K X → notin K (pref A X)
  | not_kpref: neq K M → notin K X → notin K (kpref A M X)
  | not_sum: notin K X → notin K Y → notin K (sum X Y)
  | not_par: notin K X → notin K Y → notin K (par X Y)
  | not_nu: ({a:names} notin K (X a)) → notin K (nu X)
;

%% Forward %%
LF fstep: proc → pr_lab → proc → type =
  | fs_pref: std X → fstep (pref A X) (pr_base A K) (kpref A K X)
  | fs_kpref: fstep X T X' → key T M → neq K M
           → fstep (kpref A K X) T (kpref A K X')
  | fs_suml: fstep X T X' → std Y → fstep (sum X Y) (pr_suml T) (sum X' Y)
  | fs_sumr: fstep Y T Y' → std X → fstep (sum X Y) (pr_sumr T) (sum X Y')
  | fs_parl: fstep X T X' → key T K → notin K Y
          → fstep (par X Y) (pr_parl T) (par X' Y)
  | fs_parr: fstep Y T Y' → key T K → notin K X
          → fstep (par X Y) (pr_parr T) (par X Y')
  | fs_synl: fstep X T1 X' → lab T1 (inp L) → key T1 K
          → fstep Y T2 Y' → lab T2 (out L) → key T2 K
          → fstep (par X Y) (pr_sync T1 T2) (par X' Y')
  | fs_synr: fstep X T1 X' → lab T1 (out L) → key T1 K
          → fstep Y T2 Y' → lab T2 (inp L) → key T2 K
          → fstep (par X Y) (pr_sync T1 T2) (par X' Y')
  | fs_nu: ({a:names} fstep (X a) T (X' a)) → fstep (nu X) T (nu X');
%% End %%

% Backward transitions
LF bstep: proc → pr_lab → proc → type =
  | bs_pref: std X → bstep (kpref A K X) (pr_base A K) (pref A X)
  | bs_kpref: bstep X' T X → key T M → neq K M → bstep (kpref A K X') T (kpref A K X)
  | bs_suml: bstep X' T X → std Y → bstep (sum X' Y) (pr_suml T) (sum X Y)
  | bs_sumr: bstep Y' T Y → std X → bstep (sum X Y') (pr_sumr T) (sum X Y)
  | bs_parl: bstep X' T X → key T K → notin K Y → bstep (par X' Y) (pr_parl T) (par X Y)
  | bs_parr: bstep Y' T Y → key T K → notin K X → bstep (par X Y') (pr_parr T) (par X Y)
  | bs_synl: bstep X' T1 X → lab T1 (inp L) → key T1 K → bstep Y' T2 Y → lab T2 (out L) → key T2 K → bstep (par X' Y') (pr_sync T1 T2) (par X Y)
  | bs_synr: bstep X' T1 X → lab T1 (out L) → key T1 K → bstep Y' T2 Y → lab T2 (inp L) → key T2 K → bstep (par X' Y') (pr_sync T1 T2) (par X Y)
  | bs_nu: ({a:names} bstep (X' a) T (X a)) → bstep (nu X') T (nu X)
;

%% Combined %%
LF step: proc → pr_lab → proc → type =
  | fw: fstep X T X' → step X T X'
  | bw: bstep X' T X → step X' T X;
%% End %%

%% Paths %%
LF step*: proc → proc → type =
  | id_s*: step* X X
  | st_s*: step X T Y → step* X Y
  | tr_s*: step* X Y → step* Y Z → step* X Z;
LF reachable: proc → type =
  | rch: std X → step* X Y → reachable Y;
LF conn_tr: step X T1 X' → step Y T2 Y' → type =
  | ct: {S1:step X T1 X'}{S2: step Y T2 Y'} step* X Y' → conn_tr S1 S2;
%% End %%

% In the definition of pr_lab, we admit "spurious" proof labels <T1,T2> for any T1 and T2:
% however, according to the informal definition of proof keyed labels, <T1,T2> is defined 
% only for those T1 and T2 which have the same key and complementary labels.
% As a consequence, in this encoding, there exist proof keyed labels which are not labeling
% any transition: results like realization of each proof label do not hold anymore.
% We define a new type family characterizing only the "valid" proof keyed labels.

% Type family to encode valid proof labels

%%% Connectivity, dependence and independence of proof keyed labels %%%

%% Connectivity %%
LF conn: pr_lab → pr_lab → type =
  | c_a1: conn (pr_base A K) T
  | c_a2: conn T (pr_base A K)
  | c_c1l: conn T1 T2 → conn (pr_suml T1) (pr_suml T2)
  | c_c1r: conn T1 T2 → conn (pr_sumr T1) (pr_sumr T2)
  | c_c2l: conn (pr_suml T1) (pr_sumr T2)
  | c_c2r: conn (pr_sumr T1) (pr_suml T2)
  | c_p1l: conn T1 T2 → conn (pr_parl T1) (pr_parl T2)
  | c_p1r: conn T1 T2 → conn (pr_parr T1) (pr_parr T2)
  | c_p2l: conn (pr_parl T1) (pr_parr T2)
  | c_p2r: conn (pr_parr T1) (pr_parl T2)
  | c_s1l: conn T TL → conn (pr_parl T) (pr_sync TL TR)
  | c_s1r: conn T TR → conn (pr_parr T) (pr_sync TL TR)
  | c_s2l: conn TL T → conn (pr_sync TL TR) (pr_parl T)
  | c_s2r: conn TR T → conn (pr_sync TL TR) (pr_parr T)
  | c_s3: conn T1 T1' → conn T2 T2' → conn (pr_sync T1 T2) (pr_sync T1' T2');
%% End %%

% Dependence
LF dep: pr_lab → pr_lab → type =
  | d_a1: dep (pr_base A K) T
  | d_a2: dep T (pr_base A K)
  | d_c1l: dep T1 T2 → dep (pr_suml T1) (pr_suml T2)
  | d_c1r: dep T1 T2 → dep (pr_sumr T1) (pr_sumr T2)
  | d_c2l: dep (pr_suml T1) (pr_sumr T2)
  | d_c2r: dep (pr_sumr T1) (pr_suml T2)
  | d_p1l: dep T1 T2 → dep (pr_parl T1) (pr_parl T2)
  | d_p1r: dep T1 T2 → dep (pr_parr T1) (pr_parr T2)
  | d_p2l: key T1 K → key T2 K → dep (pr_parl T1) (pr_parr T2)
  | d_p2r: key T1 K → key T2 K → dep (pr_parr T1) (pr_parl T2)
  | d_s1l: dep T TL → dep (pr_parl T) (pr_sync TL TR)
  | d_s1r: dep T TR → dep (pr_parr T) (pr_sync TL TR)
  | d_s2l: dep TL T → dep (pr_sync TL TR) (pr_parl T)
  | d_s2r: dep TR T → dep (pr_sync TL TR) (pr_parr T)
  | d_s3l: dep T1 T1' → conn T2 T2' → dep (pr_sync T1 T2) (pr_sync T1' T2')
  | d_s3r: conn T1 T1' → dep T2 T2' → dep (pr_sync T1 T2) (pr_sync T1' T2')
;

% Independence
LF indep: pr_lab → pr_lab → type =
  | i_c1l: indep T1 T2 → indep (pr_suml T1) (pr_suml T2)
  | i_c1r: indep T1 T2 → indep (pr_sumr T1) (pr_sumr T2)
  | i_p1l: indep T1 T2 → indep (pr_parl T1) (pr_parl T2)
  | i_p1r: indep T1 T2 → indep (pr_parr T1) (pr_parr T2)
  | i_p2l: key T1 K → key T2 M → neq K M → indep (pr_parl T1) (pr_parr T2)
  | i_p2r: key T1 K → key T2 M → neq K M → indep (pr_parr T1) (pr_parl T2)
  | i_s1l: indep T TL → indep (pr_parl T) (pr_sync TL TR)
  | i_s1r: indep T TR → indep (pr_parr T) (pr_sync TL TR)
  | i_s2l: indep TL T → indep (pr_sync TL TR) (pr_parl T)
  | i_s2r: indep TR T → indep (pr_sync TL TR) (pr_parr T)
  | i_s3: indep T1 T1' → indep T2 T2' → indep (pr_sync T1 T2) (pr_sync T1' T2')
;
\end{lstlisting}
\vspace{-0.5\baselineskip}
\caption{Encoding of the connectivity relation on proof labels.}
\label{fig:conn}
\end{figure}

\subsubsection{Basic properties of keys, proof labels and transitions}

Before diving into the theorems related to the causality relations, our encoding requires a small library of properties of keys, proof labels and transitions: these include the \baselink{2_basic_properties.bel}{3}{23}{decidability of equality of keys}, the \baselink{2_basic_properties.bel}{67}{125}{functionality of $\ell$ and $\kayop$}, the fact that \baselink{2_basic_properties.bel}{130}{141}{standard processes have no keys}, or the \baselink{2_basic_properties.bel}{160}{192}{loop lemma}. To provide an overview of how proofs are carried out in Beluga, we will walk through the proof of the following result: ``for all proof labels $\theta$, there exists a label $\alpha$ such that $\ell(\theta)=\alpha$''. Its code is displayed in Fig.~\ref{fig:exist-label}:

\begin{figure}[H]
\lstinputlisting[style=belugastyleframes,linerange={ExLab-End}]{code/2-basic-properties.bel}
\vspace{-0.5\baselineskip}
\caption{Proof of the existence of a label in a proof label.}
\label{fig:exist-label}
\end{figure}

The first two lines of code in Fig.~\ref{fig:exist-label} introduce a type family \type{ex_lab}, which captures the conclusions of the lemma to be proved: the type \type{ex_lab T} is inhabited whenever there exists a label \verb|A| for which \type{lab T A} holds. Defining such additional type families is the standard workaround to the lack of syntactic sugar for existentials and conjunctions in Beluga.

Thanks to the Curry-Howard isomorphism, proofs by induction are encoded through recursive functions. In Beluga, these are computation-level entities introduced by the keyword \keyw{rec}. The function \mbox{\func{existence_of_lab}} takes as input a context \verb|g| of schema \type{ctx} and a contextual object \verb|T| of type \type{pr_lab} and returns an object of type \type{ex_lab T}.
The second line of the proof asserts that the built function is total and is recursive on the second argument. These conditions are verified by Beluga's totality checker and guarantee that the function constitutes a valid proof.

The proof itself begins by introducing the argument \verb|T| through the keyword \keyw{mlam}; the other argument, the context \verb|g|, is implicit due to the use of round brackets in the function declaration. The proof proceeds by pattern matching on the object \verb|T|, which by the Curry-Howard isomorphism corresponds to case analysis on the structure of \verb|T| in the informal proof. Underscores are used to omit parameters that Beluga can infer automatically. The \verb|pr_base| and \verb|pr_sync| cases are handled immediately, since we can already provide the required object of type \type{ex_lab T}; for the remaining four cases, the proof proceeds by recursively applying the function \func{existence_of_lab} to a subterm \verb|T'| of \verb|T|, and then using the result to build the desired object. Recursive calls on structurally smaller objects correspond to applications of the inductive hypothesis in the informal proof.

We conclude this subsection by addressing the \baselink{7_complementarity.bel}{3}{76}{symmetry \emph{and} irreflexivity} of the causality relations. We report the signature of the function which proves that connectivity is symmetric:

\lstinputlisting[linerange={SymConn-End}]{code/7-complementarity.bel}

These lemmas are proved by straightforward inductions on the structure of the given predicate.

\subsection{Properties of causality relations}

Although the theorems in Section~\ref{subsec:rel-prop} are presented in a different order, here we start with the encoding of Theorem~\ref{thm:compl}, as it is straightforward and mirrors the structure of the informal proof.

\subsubsection{Encoding of Theorem~\ref{thm:compl}}

The three statements of the theorem are addressed by four recursive functions: this is because Theorem~\theoremitemref{thm:compl}{thm:compl-three} actually consists of two separate assertions, which here we prove separately. Moreover, the disjunction in the conclusions requires defining an additional type family \type{dep_or_indep}. Fig.~\ref{fig:conn-proof} displays the proof of the final assertion: ``two proof labels cannot be both dependent and independent''. We also present the signatures of the other \baselink{7_complementarity.bel}{81}{185}{\emph{recursive functions}} below.

\lstinputlisting[linerange={ComplSignatures-End}]{code/7-complementarity.bel}

\begin{figure}[ht]
\lstinputlisting[style=belugastyleframes,linerange={ComplProof-End}]{code/7-complementarity.bel}
\vspace{-0.5\baselineskip}
\caption{Proof of the statement ``two proof labels cannot be both dependent and independent''.}
\label{fig:conn-proof}
\end{figure}

The proof in Fig.~\ref{fig:conn-proof} is an example of proof by contradiction: given two objects of type \type{dep T1 T2} and \type{indep T1 T2}, the function \func{impossible_dep_and_indep} aims to derive an object of the empty type \type{false}, thereby establishing a contradiction. After introducing the arguments \verb|d| and \verb|i|, the proof proceeds by pattern matching on \verb|d|; depending on the case, the contradiction is reached in one of three distinct ways.

In case \verb|d| is built, e.g., through the constructor \verb|d_a1| (corresponding to the case A$^1$: $\alpha[k] \sdep \theta$ in the informal proof), it is immediately clear that an object \verb|i| of type \type{indep T1 T2} (i.e., $\alpha[k] \ind \theta$) does not exist: this contradiction is exhibited through the keyword \keyw{impossible}. In other subcases, such as when \verb|d| is built via \verb|d_c1l| (corresponding to C$^1_{\L}$: $+_{\L} \theta \sdep +_{\L} \theta'$, given $\theta \sdep \theta'$), the contradiction is obtained by recursively invoking \func{impossible_dep_and_indep} on smaller arguments. Finally, in the \verb|d_p2l| subcase ($\mid_{\L}\theta \sdep\ \mid_{\R}\theta'$, under the assumption $\kay{\theta} = \kay{\theta'}$), we first examine the structure of \verb|i| and find that it must have been constructed using \verb|i_p2l|. This gives us an object \verb|N| witnessing the inequality $\kay{\theta} \neq \kay{\theta'}$, which clearly contradicts our assumption; however, to complete the proof in Beluga, it is first necessary to apply the auxiliary function \func{uniqueness_of_key} for some variable unification, yielding $\kay{\theta} \neq \kay{\theta}$, followed by the function \func{irreflexive_neq}, which states the irreflexivity of the inequality of keys.

\subsubsection{Encoding of Theorem~\ref{thm:conn}}

Recall that, throughout our development, each process is assumed to be reachable. Although this hypothesis is not explicitly stated in theorems and lemmas,
it is in fact essential for proving Theorem~\ref{thm:conn} and some of its auxiliary lemmas. For this reason, before outlining its encoding, we refine its statement, making the reachability assumption explicit:

\begin{revtheorem}[Refined]
\label{thm:rev-conn} \mbox{}\\
\vspace{-4mm}
\begin{enumerate}[label=(\roman*)]
	\item If \(t_1 : X_1 \pr{fb}[\ensuremath{\scalebox{0.88}{\theta}}_1] X'_1\) and \(t_2 : X_2 \pr{fb}[\ensuremath{\scalebox{0.88}{\theta}}_2] X'_2\) are connected and $X_1$ is reachable\footnote{The reachability of the only $X_1$ is enough to deduce the reachability of any other process in the statement, given the existence of a path from $X_1$ to such processes.}, then \(\theta_1 \conn \theta_2\). \label{thm:rev-conn-one}
	\item If \(\theta_1 \conn \theta_2\), then there exist \(t_1 : X_1 \pr{fb}[\ensuremath{\scalebox{0.88}{\theta}}_1] X'_1\) and \(t_2 : X_2 \pr{fb}[\ensuremath{\scalebox{0.88}{\theta}}_2] X'_2\), with $X_1$ reachable, such that \(t_1\) and \(t_2\) are connected.  \label{thm:rev-conn-two}
\end{enumerate}
\end{revtheorem}

Since the two statements are encoded by two distinct functions, we discuss them separately. The proof of Theorem~\theoremitemref{thm:conn}{thm:rev-conn-one} is given by the following function \linkfun{4_connectivity_relationship_one.bel}{3}{291}{conn_rel_one}:

\lstinputlisting[linerange={StatementOne-End}]{code/4-prop-4-4-one.bel}

The function takes as inputs two transitions \verb|S1| and \verb|S2|, whose typing judgments introduce the names of each involved parameter, such as the process \verb|X1|; these are followed by the further assumptions of reachability of \verb|X1| and connectivity of \verb|S1| and \verb|S2|. The function returns a derivation of the connectivity of the proof labels \verb|T1| and \verb|T2|.

Although our encoding may appear different -- and somewhat longer -- than the proof presented in \cite{aubert24}, it is, in essence, faithful to the same underlying structure. The original proof leverages Lemma~\ref{lem:connected-origin} to establish the equality of the processes $O_{X_1}$ and $O_{X_2}$, the origins of the sources of the connected transitions $t_1 : X_1 \pr{fb}[\ensuremath{\scalebox{0.88}{\theta}}_1] X'_1$ and $t_2 : X_2 \pr{fb}[\ensuremath{\scalebox{0.88}{\theta}}_2] X'_2$. It proceeds by induction on $O_{X_1}$, observing that its structure determines that of the processes and transitions in the same environment (e.g., if the outermost operator of $O_{X_1}$ is a sum, the same applies to $X_1$). The proof then concludes either directly or by applying the induction hypothesis to transitions involving specific subprocesses.

Below, we outline the changes and technical considerations brought by our encoding of this argument:

\begin{itemize}
\item The formalized proof proceeds by pattern matching on an object \verb|D| of type \type{std OX1}, rather than directly on the process \verb|OX1|. This is essentially equivalent, since the type family \type{std proc}, which asserts that a process is standard, is itself defined by pattern matching on the underlying process.
\item It is not necessary to encode Lemma~\ref{lem:connected-origin}. The reachability of $X_1$, together with the existence of a path from $X_1$ to $X_2$, provides us a path between $O_{X_1}$ and $X_2$; this path is enough to determine the structure of $X_2$, known the structure of $O_{X_1}$.
\item The proof requires analyzing the structure of the given transitions \verb|S1| and \verb|S2|. Since combined transitions are either forward or backward, and each have their own constructors, this results in four levels of nested pattern matching. While most of the subcases can be unified in the informal proof, Beluga requires them to be treated separately: this is the primary reason for the proof's length. To improve efficiency, certain assertions have been moved earlier in the proof tree compared to their position in the informal version.
\item The informal proof takes for granted structural properties such as: ``given a path whose source is a sum process, the target is also a sum process'', or ``given a path between two sum processes, there exists a path between their left addends''. In the encoding, these results must be explicitly stated and proved, resulting in 16 \baselink{3_lemmas_connectivity_relationship_one.bel}{1}{289}{\emph{additional lemmas}}. Some of these require classical techniques such as mutual recursion or strengthening of contextual judgments, which are described in \cite{momigliano24}. We report the signatures of two of these functions:
\lstinputlisting[linerange={AddLemmaOne-End,AddLemmaTwo-End}]{code/3-lemmas-prop-4-4-one.bel}
\end{itemize}

The formalization of Theorem~\theoremitemref{thm:conn}{thm:rev-conn-two} requires encoding Lemma~\ref{lem:realis}, which states that every proof label $\theta$ is realised by some process $r(\theta)$ -- that is, there exist processes $X_1$, $X_2$ and $r(\theta)$ such that $r(\theta) \pr{fb}^* X_1 \pr{fb}[\ensuremath{\scalebox{0.88}{\theta}}] X_2$. The original proof in \cite{aubert24}, however, goes further: it builds a process $r(\theta)$ which is standard and directly performs a single forward transition $r(\theta) \pr{f}[\ensuremath{\scalebox{0.88}{\theta}}] X_2$ (in other words, $r(\theta)$ and $X_1$ coincide). Our encoding reflects this stronger formulation by specializing the original Definition~\ref{def:real} with the following type family \type{realised}:

\begin{lstlisting}[linerange={Realized-End}]
%%% Auxiliary lemmas for Proposition 4.4 (second part)

% Type family encoding realization of a proof label
% Unlike the paper, we provide a slightly stronger definition. Proposition B.12 still holds.
%% Realized %%
LF realised: pr_lab → type =
  | rl: std X → fstep X T X' → realised T;
%% End %%

% Proposition B.12: every valid proof label is realized
%% RealLemma %%
rec pr_lab_is_realised: (g:ctx) [g ⊢ valid T] → [g ⊢ realised T] = ...
%% End %%
/ total g (prop_b12 _ _ g) /
fn d ⇒ case d of
  | [g ⊢ v_base] ⇒ [g ⊢ rl (std_pref std_null) (fs_pref std_null)]
  | [g ⊢ v_suml V] ⇒ let [g ⊢ rl D F] = prop_b12 [g ⊢ V] in [g ⊢ rl (std_sum D std_null) (fs_suml F std_null)]
  | [g ⊢ v_sumr V] ⇒ let [g ⊢ rl D F] = prop_b12 [g ⊢ V] in [g ⊢ rl (std_sum std_null D) (fs_sumr F std_null)]
  | [g ⊢ v_parl V] ⇒ let [g ⊢ rl D F]:[g ⊢ realized T] = prop_b12 [g ⊢ V] in
    let [g ⊢ ex_k H] = existence_of_key [g ⊢ T] in
    [g ⊢ rl (std_par D std_null) (fs_parl F H not_null)]
  | [g ⊢ v_parr V] ⇒ let [g ⊢ rl D F]:[g ⊢ realized T] = prop_b12 [g ⊢ V] in
    let [g ⊢ ex_k H] = existence_of_key [g ⊢ T] in
    [g ⊢ rl (std_par std_null D) (fs_parr F H not_null)]
  | [g ⊢ v_synl V1 V2 L1 H1 L2 H2] ⇒ let [g ⊢ rl D1 F1] = prop_b12 [g ⊢ V1] in
    let [g ⊢ rl D2 F2] = prop_b12 [g ⊢ V2] in
    [g ⊢ rl (std_par D1 D2) (fs_synl F1 L1 H1 F2 L2 H2)]
  | [g ⊢ v_synr V1 V2 L1 H1 L2 H2] ⇒ let [g ⊢ rl D1 F1] = prop_b12 [g ⊢ V1] in
    let [g ⊢ rl D2 F2] = prop_b12 [g ⊢ V2] in
    [g ⊢ rl (std_par D1 D2) (fs_synr F1 L1 H1 F2 L2 H2)]
;

% The following lemmas 

% Given a path between X and X', there is a path between X+0 and X'+0
rec step*_sumleft_null: (g:ctx) [g ⊢ step* X X'] → [g ⊢ step* (sum X null) (sum X' null)] =
/ total s (step*_sumleft_null _ _ _ s) /
fn s ⇒ case s of
  | [g ⊢ id_s*] ⇒ [g ⊢ id_s*]
  | [g ⊢ st_s* S] ⇒ (case [g ⊢ S] of
       | [g ⊢ fw F] ⇒ [g ⊢ st_s* (fw (fs_suml F std_null))]
       | [g ⊢ bw B] ⇒ [g ⊢ st_s* (bw (bs_suml B std_null))])
  | [g ⊢ tr_s* S1* S2*] ⇒ let [g ⊢ S1*'] = step*_sumleft_null [g ⊢ S1*] in
    let [g ⊢ S2*'] = step*_sumleft_null [g ⊢ S2*] in [g ⊢ tr_s* S1*' S2*']
;

% Given a path between X and X', there is a path between 0+X and 0+X'
rec step*_sumright_null: (g:ctx) [g ⊢ step* X X'] → [g ⊢ step* (sum null X) (sum null X')] =
/ total s (step*_sumright_null _ _ _ s) /
fn s ⇒ case s of
  | [g ⊢ id_s*] ⇒ [g ⊢ id_s*]
  | [g ⊢ st_s* S] ⇒ (case [g ⊢ S] of
       | [g ⊢ fw F] ⇒ [g ⊢ st_s* (fw (fs_sumr F std_null))]
       | [g ⊢ bw B] ⇒ [g ⊢ st_s* (bw (bs_sumr B std_null))])
  | [g ⊢ tr_s* S1* S2*] ⇒ let [g ⊢ S1*'] = step*_sumright_null [g ⊢ S1*] in
    let [g ⊢ S2*'] = step*_sumright_null [g ⊢ S2*] in [g ⊢ tr_s* S1*' S2*']
;

% Given a path between X and X' and a standard process Y, there is a path between X|Y and X'|Y
rec step*_parleft_std: (g:ctx) [g ⊢ step* X X'] → [g ⊢ std Y] → [g ⊢ step* (par X Y) (par X' Y)] =
/ total s (step*_parleft_std _ _ _ _ s _) /
fn s,d ⇒ case s of
  | [g ⊢ id_s*] ⇒ [g ⊢ id_s*]
  | [g ⊢ st_s* S] ⇒ (case [g ⊢ S] of
       | [g ⊢ fw F]:[g ⊢ step _ T _] ⇒ let [g ⊢ D] = d in
         let [g ⊢ ex_kn _ H N] = no_key_of_prlab_in_std [g ⊢ T] [g ⊢ D] in [g ⊢ st_s* (fw (fs_parl F H N))]
       | [g ⊢ bw B]:[g ⊢ step _ T _] ⇒ let [g ⊢ D] = d in
         let [g ⊢ ex_kn _ H N] = no_key_of_prlab_in_std [g ⊢ T] [g ⊢ D] in [g ⊢ st_s* (bw (bs_parl B H N))])
  | [g ⊢ tr_s* S1* S2*] ⇒ let [g ⊢ S1*'] = step*_parleft_std [g ⊢ S1*] d in
    let [g ⊢ S2*'] = step*_parleft_std [g ⊢ S2*] d in [g ⊢ tr_s* S1*' S2*']
;

% Given a transition between X and X' and a standard process Y, there is a transition between Y|X and Y|X'
rec step*_parright_std: (g:ctx) [g ⊢ step* X X'] → [g ⊢ std Y] → [g ⊢ step* (par Y X) (par Y X')] =
/ total s (step*_parright_std _ _ _ _ s _) /
fn s,d ⇒ case s of
  | [g ⊢ id_s*] ⇒ [g ⊢ id_s*]
  | [g ⊢ st_s* S] ⇒ (case [g ⊢ S] of
       | [g ⊢ fw F]:[g ⊢ step _ T _] ⇒ let [g ⊢ D] = d in
         let [g ⊢ ex_kn _ H N] = no_key_of_prlab_in_std [g ⊢ T] [g ⊢ D] in [g ⊢ st_s* (fw (fs_parr F H N))]
       | [g ⊢ bw B]:[g ⊢ step _ T _] ⇒ let [g ⊢ D] = d in
         let [g ⊢ ex_kn _ H N] = no_key_of_prlab_in_std [g ⊢ T] [g ⊢ D] in [g ⊢ st_s* (bw (bs_parr B H N))])
  | [g ⊢ tr_s* S1* S2*] ⇒ let [g ⊢ S1*'] = step*_parright_std [g ⊢ S1*] d in
    let [g ⊢ S2*'] = step*_parright_std [g ⊢ S2*] d in [g ⊢ tr_s* S1*' S2*']
;

% If a standard process makes a transition F indexed by the key K,
% then any key different than K does not occur in the target of F
rec lemma_fstep_std: (g:ctx) [g ⊢ std X] → [g ⊢ fstep X T Y] →
  [g ⊢ key T M[]] → [g ⊢ neq M[] K[]] → [g ⊢ notin K[] Y] =
/ total d (lemma_fstep_std _ _ _ _ _ _ d _ _ _) /
fn d,f,h,i ⇒ let [g ⊢ I]:[g ⊢ neq _ K[]] = i in
case d of
  | [g ⊢ std_null] ⇒ impossible f
  | [g ⊢ std_pref D] ⇒ let [g ⊢ fs_pref _] = f in
    let [g ⊢ refk] = uniqueness_of_key h [g ⊢ key_base] in
    let [g ⊢ N] = no_key_in_std [⊢ K] [g ⊢ D] in
    let [g ⊢ I] = symmetric_neq i in [g ⊢ not_kpref I N]
  | [g ⊢ std_sum D1 D2] ⇒ (case f of
       | [g ⊢ fs_suml FL _] ⇒ let [g ⊢ key_suml H] = h in
         let [g ⊢ N1] = lemma_fstep_std [g ⊢ D1] [g ⊢ FL] [g ⊢ H] i in
         let [g ⊢ N2] = no_key_in_std [⊢ K] [g ⊢ D2] in [g ⊢ not_sum N1 N2]
       | [g ⊢ fs_sumr FR _] ⇒ let [g ⊢ key_sumr H] = h in
         let [g ⊢ N1] = no_key_in_std [⊢ K] [g ⊢ D1] in
         let [g ⊢ N2] = lemma_fstep_std [g ⊢ D2] [g ⊢ FR] [g ⊢ H] i in [g ⊢ not_sum N1 N2])
  | [g ⊢ std_par D1 D2] ⇒ (case f of
       | [g ⊢ fs_parl F1 _ _] ⇒ let [g ⊢ key_parl H] = h in
         let [g ⊢ N1] = lemma_fstep_std [g ⊢ D1] [g ⊢ F1] [g ⊢ H] i in
         let [g ⊢ N2] = no_key_in_std [⊢ K] [g ⊢ D2] in [g ⊢ not_par N1 N2]
       | [g ⊢ fs_parr F2 _ _] ⇒ let [g ⊢ key_parr H] = h in
         let [g ⊢ N1] = no_key_in_std [⊢ K] [g ⊢ D1] in
         let [g ⊢ N2] = lemma_fstep_std [g ⊢ D2] [g ⊢ F2] [g ⊢ H] i in [g ⊢ not_par N1 N2]
       | [g ⊢ fs_synl F1 _ H1 F2 _ H2] ⇒  let [g ⊢ key_sync H] = h in
         let [g ⊢ refk] = uniqueness_of_key [g ⊢ H] [g ⊢ H1] in
         let [g ⊢ N1] = lemma_fstep_std [g ⊢ D1] [g ⊢ F1] [g ⊢ H1] i in
         let [g ⊢ N2] = lemma_fstep_std [g ⊢ D2] [g ⊢ F2] [g ⊢ H2] i in [g ⊢ not_par N1 N2]
       | [g ⊢ fs_synr F1 _ H1 F2 _ H2] ⇒ let [g ⊢ key_sync H] = h in
         let [g ⊢ refk] = uniqueness_of_key [g ⊢ H] [g ⊢ H1] in
         let [g ⊢ N1] = lemma_fstep_std [g ⊢ D1] [g ⊢ F1] [g ⊢ H1] i in
         let [g ⊢ N2] = lemma_fstep_std [g ⊢ D2] [g ⊢ F2] [g ⊢ H2] i in [g ⊢ not_par N1 N2])
  | [g ⊢ std_nu \a.D] ⇒ let [g ⊢ fs_nu \a.F] = f in
    let [g ⊢ H] = h in
    let [g,a:names ⊢ N] = lemma_fstep_std [g,a:names ⊢ D] [g,a:names ⊢ F] [g,a:names ⊢ H[..]] [g,a:names ⊢ I[..]] in [g ⊢ not_nu \a.N]
;
\end{lstlisting}
For any proof label \verb|T|, \type{realised T} is non empty iff \type{std X} and \type{fstep X T X'} hold for some \verb|X| and \verb|X'|. The following recursive function \linkfun{5_lemmas_connectivity_relationship_two.bel}{10}{29}{pr_lab_is_realised} encodes the proof of Lemma~\ref{lem:realis}:
\lstinputlisting[linerange={RealLemma-End}]{code/5-lemmas-prop-4-4-two.bel}
The proof is a straightforward induction on the structure of the assumption \type{valid T}.

Other than relying on Lemma~\ref{lem:realis}, the proof of Theorem~\theoremitemref{thm:conn}{thm:rev-conn-two} in \cite{aubert24} assumes auxiliary results such as the following: ``if $O_{X}$ realises $X$ and $O_{Y}$ realises $Y$, then $O_{X} \mid O_{Y}$ realises $X \mid Y$''. While this result holds in the particular context of Theorem~\theoremitemref{thm:conn}{thm:rev-conn-two}, where $X \mid Y$ is known to be reachable and is able to perform a synchronization, it does not hold in general. For instance, consider $X_1=a[k]$ and $X_2=b[k]$: the parallel composition $a[k] \mid b[k]$ is not reachable from $a \mid b$. Moreover, even when such conditions are met, building a constructive proof is far from straightforward. These issues led us to revisit the entire argument and develop the following proof strategy for Theorem~\theoremitemref{thm:conn}{thm:rev-conn-two}:
\begin{enumerate}
\item First, we consider the case where neither $\ell(\theta_1)$ nor $\ell(\theta_2)$ is $\tau$ and prove that the diagram in Fig.~\ref{fig:strat-one} holds. The hypothesis excludes the cases in which $\theta_1$ and $\theta_2$ label synchronizations, thus ruling out the scenarios in which the aforementioned auxiliary lemma occurs. The proved result goes beyond establishing the connectivity of two combined transitions labelled by $\theta_1$ and $\theta_2$: both transitions are forward, and the processes $X_1$ and $X_2'$ are either identical or connected by a single combined transition. Additionally, we show that at least one among $X_1$ and $X_2'$ is standard.\label{thm:case-one}

\item We then move to the general case, proving that for any connected pair of proof labels $\theta_1$ and $\theta_2$ the diagram in Fig.~\ref{fig:strat-two} holds. Analogously to the previous point, the transitions labelled by $\theta_1$ and $\theta_2$ are forward, meaning that our statement is slightly more specific than the original formulation of Theorem~\theoremitemref{thm:conn}{thm:rev-conn-two}. This refinement helps eliminating non-existent subcases that would arise in the nested pattern matching of combined transitions.
\end{enumerate}

\begin{figure}[H]
  \centering
  \fbox{%
  \begin{subfigure}[b]{0.45\textwidth}
    \centering
    \begin{tikzpicture}[
        x={(1.5, 0)},
        y={(0, 1.1)},
        baseline,
        anchor=base
      ]
      \node (x1) at (0, 0){\(X_1\)};
      \node (x2) at (2.4, 1.2){\(X_2\)};
      \node (x1p) at (0, -1.2){\(X_1'\)};
      \node (x2p) at (2.4, 0){\(X_2'\)};
      \node (mid) at (1.2, 0){$\pr{fb}[\qquad \qquad \qquad \quad]$};

      \draw[{Bar[]}->] (x1) -- node[lbl, xshift=-10, yshift=-3]{\(\theta_1\)} (x1p);
      \draw[{Bar[]}->] (x2p) -- node[lbl, xshift=10, yshift=-3]{\(\theta_2\)} (x2);
    \end{tikzpicture}
    \caption{Base case of the proof strategy.}
    \label{fig:strat-one}
  \end{subfigure}
  \hspace{0.05\textwidth}
  \begin{subfigure}[b]{0.45\textwidth}
    \centering
    \begin{tikzpicture}[
	x={(1.5, 0)},
	y={(0, 1.1)},
	baseline,
	anchor=base
	]
	\node (x1) at (0, 0){\(X_1\)};
	\node (x2) at (2.4, 1.2){\(X_2\)};
	\node (x1p) at (0, -1.2){\(X_1'\)};
	\node (x2p) at (2.4, 0){\(X_2'\)};
	\node (mid1) at (0.6, 0){$\pr{fb}[\quad \ \ \ ]$};
	\node (mid2) at (1.2, 0){$\cdots$};
	\node (mid3) at (1.8, 0){$\pr{fb}[\quad \ \ \ ]$};

	\draw[{Bar[]}->] (x1) -- node[lbl, xshift=-10, yshift=-3]{\(\theta_1\)} (x1p);
	\draw[{Bar[]}->] (x2p) -- node[lbl, xshift=10, yshift=-3]{\(\theta_2\)} (x2);
    \end{tikzpicture}
    \caption{General case of the proof strategy.}
    \label{fig:strat-two}
  \end{subfigure}
  }
  \caption{Proof strategy for Theorem~\theoremitemref{thm:conn}{thm:rev-conn-two}.}
  \label{fig:main-fig}
\end{figure}
	
	In the general case, when $\theta_1$ and $\theta_2$ label synchronizations (e.g., when $\theta_1=\langle|_{\L}\theta^1_L,\,|_{\R}\theta^1_R\rangle$), the labels of their subterms (e.g., $\theta^1_L$ and $\theta^1_R$) are not $\tau$: this detail allows us to apply the base case of the proof strategy, which provides richer information than the inductive hypothesis of the general Theorem~\theoremitemref{thm:conn}{thm:rev-conn-two}. That additional information is essential: it enables us to build the desired path between $X_1$ and $X_2'$, actually with at most two transition steps.

The encoding of Theorem~\theoremitemref{thm:conn}{thm:rev-conn-two} follows the plan outlined. The base case makes use of the following elements: a type family \type{lab_not_tau}, characterizing proof keyed labels whose label is not $\tau$; a type family {\type{max_one_step}, encoding the conclusions of the statement; and the recursive function \mbox{\linkfun{6_connectivity_relationship_two.bel}{24}{196}{conn_rel_two_base}}, which proves it.
\lstinputlisting[linerange={ThmTwoBase-End}]{code/6-prop-4-4-two.bel}
The proof of this result is given by a long induction on the structure of the given connectivity relation. The predicates (\verb|lab Ti Lj|), for \verb|i|,\verb|j| in $\{1,2,3\}$, which occur in the type family \type{max_one_step}, are a technical detail which helps completing few subcases of the proof. 

Next, the general case of the proof is addressed by the recursive function \mbox{\linkfun{6_connectivity_relationship_two.bel}{205}{581}{conn_rel_two_fstep}} below, which relies on a dedicated type family as well:
\lstinputlisting[linerange={ThmTwoFstep-End}]{code/6-prop-4-4-two.bel}
The proof is given by a long induction on the structure of the given connectivity relation. It requires encoding \baselink{5_lemmas_connectivity_relationship_two.bel}{32}{124}{\emph{auxiliary lemmas}} such as the following: ``given a path between two processes $X$ and $X'$, there is a path between $X+\textbf{0}$ and $X'+\textbf{0}$'', or: ``given a forward transition $X \pr{f}[\ensuremath{\scalebox{0.88}{\theta}}] X'$ where $X$ is standard and $\kayop(\theta)=k$, then any key $m \neq k$ does not occur in $X'$''.

Finally, Theorem~\theoremitemref{thm:conn}{thm:rev-conn-two} is encoded by the following function \func{conn_rel_two}. It calls the function \func{conn_rel_two_fstep}, applies the loop lemma to reverse one of the two forward transitions, and has all the ingredients to conclude:
\lstinputlisting[linerange={ThmTwo-End}]{code/6-prop-4-4-two.bel}

%% file: sections/conclusions.tex
\section{Conclusions and Future Work}\label{sec:conclusions}

We begin with a brief technical overview of the encoding. The complete formalization consists of less than 2000 lines of code and includes a total of 49 theorems and lemmas. Among them, 13 are direct translations of results stated in Section~\ref{sec:ccskp}, while the remaining 36 are technical and auxiliary lemmas introduced to support the encoding.

Beluga has proved to be a reliable and expressive proof assistant, well-suited to represent the definitions and properties of \pccsk. 
Its use of higher-order abstract syntax (HOAS) offers a convenient approach to handling restrictions -- even though \pccsk does not feature a particularly complex binding structure, unlike, for instance, the $\pi$-calculus. Furthermore, Beluga's explicit proof style provides a transparency that is often lost in proof assistants that rely heavily on automation.

However, the lack of automation also comes with drawbacks, mainly the increased length of proof terms. This also follows from the lack of syntactic sugar for existentials, conjunctions and disjunctions, which leads to defining additional type families or splitting theorem statements. Additionally, Beluga provides no built-in mechanism to simplify repeated proof patterns, requiring each similar subcase to be handled individually.

Whether the overall outcome is favorable depends largely on the specific system one aims to formalize. For languages with rich binding structures, the benefits of HOAS alone may outweigh the trade-offs. In our case, however, this advantage is less significant, and we believe that other proof assistants (such as Rocq~\cite{BertotC04}) might be a better fit for formalizing the system at hand.

To the best of our knowledge, this work provides the first formalization of a reversible concurrent calculus in a proof assistant. We have formally verified the correctness of the notions and results presented in Section~\ref{sec:ccskp}. We gained a deeper understanding of the system itself, leading to refinements in both definitions and proofs; in particular, we provided an alternative way to represent proof labels compared to the informal definition.

This work lays the foundations for future reversible concurrent calculi formalizations. The encoding can be adapted to cover the subsystems of \pccsk, i.e., \ccs and \ccsk, and can be mapped to existing \ccs formalizations. Moreover, it could be extended to include additional portions of \cite{aubert25}. Additionally, it could be translated into other proof assistants, such as Rocq, which are potentially better suited for representing this reversible process calculus. Finally, it can serve as a reference point for future formalizations of other reversible concurrent calculi, such as \rccs.